\documentclass[shortpaper,twoside,web]{ieeetran}

\usepackage{textcomp}

\usepackage{algorithm}
\usepackage{algpseudocode}

\usepackage{amssymb}
\usepackage{amsmath}
\usepackage{amsfonts}                                
\usepackage[latin1] {inputenc}
\usepackage{enumerate}
\usepackage{mathrsfs}
\usepackage{tabulary}
\usepackage{cite}
\usepackage{psfrag}
\usepackage{graphicx}         
\usepackage[dvipsnames]{xcolor} 
\usepackage{epsfig} 
\usepackage{epstopdf}
\usepackage{times} 
\def\qedp{\hspace*{\fill}~{\tiny $\blacksquare$}}
\newtheorem{theorem}{Theorem}
\newtheorem{itlemma}{Lemma}
\newtheorem{itdefinition}{Definition}
\newtheorem{itproposition}{Proposition}
\newtheorem{itresult}{Result}
\newtheorem{itremark}{Remark}
\newtheorem{itassumption}{Assumption}
\newtheorem{itcorollary}{Corollary}
\newtheorem{itexample}{Example}
\newenvironment{definition}{\begin{itdefinition}\rm}{\end{itdefinition}}

\newenvironment{remark}{\begin{itremark}\rm}{\end{itremark}}
\newenvironment{assumption}{\begin{itassumption}\rm}{\end{itassumption}}
\newenvironment{lemma}{\begin{itlemma}\rm}{\end{itlemma}}

\AtBeginDocument{
	\addtolength{\abovedisplayskip}{-0.9ex}
	\addtolength{\abovedisplayshortskip}{-0.9ex}
	\addtolength{\belowdisplayskip}{-0.9ex}
	\addtolength{\belowdisplayshortskip}{-0.9ex}
}

\title{\LARGE Quantized distributed Nash equilibrium seeking under DoS attacks} 

\author{
	Shuai Feng, \emph{Member, IEEE}, Maojiao Ye, \emph{Senior Member, IEEE},\\ Lihua Xie, \emph{Fellow, IEEE}, Shengyuan Xu, \emph{Senior Member, IEEE}
	\thanks{Shuai Feng, Maojiao Ye and Shengyuan Xu are with the School of Automation, Nanjing University of Science and Technology, Nanjing 210094, China ({\tt\small s.feng@njust.edu.cn}, {\tt\small mjye@njust.edu.cn}, {\tt\small			syxu@njust.edu.cn}). 
		Lihua Xie is with the School of Electrical and Electronics Engineering, Nanyang Technological University, Singapore 68000, Singapore ({\tt \small ELHXIE@ntu.edu.sg}).} 
}

\begin{document}
	\maketitle
	\begin{abstract}
		This paper studies distributed Nash equilibrium (NE) seeking under Denial-of-Service (DoS) attacks and quantization. The players can only exchange information with their own direct neighbors. The transmitted information is subject to quantization and packet losses induced by malicious DoS attacks. We propose a quantized distributed NE seeking strategy based on the approach of dynamic quantized consensus. To solve the quantizer saturation problem caused by DoS attacks, the quantization mechanism is equipped to have zooming-in and holding capabilities, in which the holding capability is consistent with the results in quantized consensus under DoS. 
		A sufficient condition on the number of quantizer levels is provided, under which the quantizers are free from saturation under DoS attacks. The proposed distributed quantized NE seeking strategy is shown to have the so-called maximum resilience to DoS attacks. Namely, if the bound characterizing the maximum resilience is violated, an attacker can deny all the transmissions and hence distributed NE seeking is impossible.  
	\end{abstract}

\section{Introduction}
Game theory has attracted tremendous attention owing to its advances in decision making. It has great potentials in the applications such as wireless communication networks, defending cyber attacks and cloud computing \cite{osborne1994course, zhu2015game}. Nash equilibrium (NE) seeking is one of the most important problems in non-cooperative games. Recently, distributed NE seeking with partial-decision information becomes a new thriving direction, in which players do not need to have direct access to the information of all the other players \cite{salehisadaghiani2019distributed,xu2022hybrid}.

For distributed NE seeking with partial-decision information, exchanging information among players via communication networks is necessary, by which the players can estimate the actions of the other players \cite{ye2017distributed}. It is well known that state estimation significantly relies on the underlying communication networks. However, communication networks are not always perfect, and are prone to have problems such as delay, packet losses, bandwidth limitation and cyber attacks\cite{cardenas2009challenges}. A distributed NE seeking game may fail in the presence of the above problems.    

Under limited bandwidth, signals can be subject
to coarse quantization \cite{ishii2002limited}. The consequence of quantization
on control/measurement signals needs to be taken into account
at the design stage. Otherwise, when quantizer saturation occurs, a closed-loop control system can be unstable. Quantized control has been widely studied in the literature (see the seminal papers \cite{867021, 1310461} and their citations). Recently, there are a few works investigating distributed NE seeking \cite{9878291,chen2022distributed,nekouei2016performance} and distributed optimization \cite{doan2020convergence,zhang2018distributed,xiong2022quantized,yi2014quantized} under quantization. In \cite{9878291}, the authors developed distributed NE seeking algorithms under uniform and logarithmic quantization. However, the quantizer ranges preventing saturation were not characterized. The work \cite{chen2022distributed} provided a sufficient condition of quantization range, under which the actions of players exponentially converge to a NE and the quantizer is not saturated. The early work \cite{nekouei2016performance} investigated distributed quantized NE seeking, in which the graph of the communication topology is a complete graph.

Under an unreliable network with cyber attacks, players' actions may not reach a NE \cite{wang2020distributed}. Our paper considers one of the most common cyber attacks: Denial-of-Service (DoS) attacks. The intention of DoS attackers is to maliciously jam the networks. Communication failures induced by DoS can exhibit a temporal profile quite different from the one induced by genuine packet losses.
Particularly, packet dropouts induced by DoS need not follow a given class of probability distributions \cite{sastry}. Therefore, the techniques relying on probabilistic arguments for random packet losses may not be applicable in the context of DoS attacks. In the past decade, control over DoS-corrupted networks has been widely studied \cite{liu2022data,deng2022resilient}. The paper \cite{deng2022resilient} proposed a new distributed resilient control method to address the output regulation problem under DoS attacks, without pre-knowing the dynamics of the leader.
Since DoS attacks can induce communication failures, the existing distributed NE seeking strategies and results, e.g., in \cite{9878291, ye2017distributed, chen2022distributed,rao2023distributed}, may not always hold.

When quantization and DoS attacks coexist, NE seeking problems are challenging because one needs to ensure that quantizers should not saturate under DoS attacks. For example, under DoS attacks, states could diverge. If a quantization scheme has only zooming-in capabilities as in \cite{chen2022distributed}, quantizer saturation will occur under DoS attacks. To mitigate quantizer saturation problems in the presence of random or malicious packet losses, one can develop quantization strategies with both zooming-in and zooming-out capabilities, for centralized systems \cite{you2010minimum} and multi-agent systems \cite{feng2023tcns}. For instance, the work \cite{you2010minimum} developed zooming-in and out quantization strategies to stabilize centralized systems under random packet losses. In \cite{feng2023tcns}, the authors proposed a zooming-in and out strategy to tackle quantized consensus problems under DoS attacks. The approach in our paper is inspired by the ideas of consensus based NE seeking in \cite{ye2017distributed}, dynamic quantized consensus under DoS attacks in \cite{feng2023tcns} and quantized consensus in the seminal papers \cite{you2011network, li2010distributed}.

Our paper deals with quantized distributed NE seeking problems under DoS attacks. We are interested in three questions: 1. how to design a new quantization scheme realizing NE seeking and preventing quantizer saturation under DoS attacks; 2. how robust is the quantized distributed NE seeking algorithm; 3. under DoS attacks, what is the lower bound of the number of quantization levels preventing quantizer saturation. Different from the Lyapunov-based approaches \cite{9878291, chen2022distributed}, we develop a new path to answer the above questions, which is aligned with the approaches for studying quantized consensus \cite{you2011network, li2010distributed, feng2023tcns}. We mention that such a new development is challenging from the technical viewpoint because NE seeking problems involve nonlinear functions, under which some key techniques for linear-system consensus in \cite{you2011network, li2010distributed, feng2023tcns} are not applicable. The main contributions of this paper are listed as follows:     
\begin{itemize}
	\item We develop a new quantized distributed NE seeking strategy with zooming-in and holding capabilities, which can tackle quantizer saturation problems in the presence of DoS attacks. The NE seeking strategy reaches the so-called ``the maximum resilience" against DoS attacks as in \cite{feng2017resilient}.  
	\item The lower bounds of the number of quantizer levels are explicitly provided, under which the quantizers are not saturated even under DoS attacks.   
\end{itemize}
Specifically, regarding the first contribution, if the bound characterizing ``the maximum resilience" is violated, an attacker can deny all the transmissions \cite{feng2017resilient} and consequently distributed NE seeking is impossible. Detailed comparisons with relevant papers are presented in Remark \ref{remark 5}.

This paper is organized as follows. In Section II, we introduce the framework consisting of the dynamics of players, the class of DoS attacks, and the control objectives. Section III presents the quantized distributed NE seeking strategy with zooming-in and holding capabilities. Main results are presented in Section IV, including lower bounds of quantization levels preventing quantizer saturation and a bound characterizing the resilience.  
A numerical example is presented in Section V, and finally Section VI ends the paper with conclusions and future research.

\textbf{Notation.} We denote by  $\mathbb R$ the set of reals. Given $b \in \mathbb R$, $\mathbb R_{\geq b}$ and $\mathbb R_{>b}$ denote the sets of reals no smaller than $b$ and reals greater than $b$, respectively; $\mathbb R_{\le b}$ and $\mathbb R_{<b}$ represent the sets of reals no larger than $b$ and reals smaller than $b$, respectively; $\mathbb Z$ denotes the set of integers. For any $c \in \mathbb Z$, we denote $\mathbb Z_{\ge c} := \{c,c+ 1,\cdots\}$. 
Given a vector $\zeta$ and a matrix $\Omega$, let $\|\zeta\|$ and $\|\zeta\|_\infty$ denote the $2 $- and $\infty$-norms of vector $\zeta$, respectively,  
and $\|\Omega\|$ and $\|\Omega\|_\infty$ represent the corresponding induced norms of matrix $\Omega$. Moreover, $\rho(\Omega)$ denotes the spectral radius of $\Omega$. Given an interval $\mathcal{I}$, $|\mathcal{I}|$ denotes its length. The Kronecker product is denoted by $\otimes$. 
For $\omega \in \mathbb Z _{\ge 2}$, let $I _\omega$ denote the identity matrix with dimension $\omega$. Let $\mathbf 1$ denote a vector $[1 \cdots 1 ]^T$ with a compatible dimension.

\section{Framework}

\emph{Communication graph:}
We let graph $\mathcal{G}$ denote the communication topology among the $N$ players, which are indexed by $1, 2, \cdots, N$. Let $\mathcal{N}=\{1, 2, \cdots, N \}$ be the set of players and let $\mathcal N _ i $ denote the set of the neighbor players of player $i$. In this paper, we assume that the graph $\mathcal{G}$ is undirected and connected. Let $\mathcal A = [a_{ij}] \in \mathbb{R}^{N\times N} $ ($i,j \in \mathcal{V}$) denote the adjacency matrix of the graph $\mathcal {G}$, where $a_{ij} > 0$ if and only if $j \in \mathcal N_i$ and $a_{ii}=0$. Define the Laplacian matrix $\mathcal{L} = [l_{ij}]  \in \mathbb{R}^{N\times N} $, in which $l_{ii} = \sum_{j = 1 }^{N} a_{ij}$ and $l_{ij} = - a_{ij} $ if $i \ne j $.

\begin{definition}
	(\emph{A normal form game}): A normal form game $\Gamma$ is outlined as $\Gamma : = (\mathcal N, X, f)$ with $X = X_1 \times  \cdots \times X_N$ and $f = (f_1(  x), \cdots, f_N(  x))$, in which $X_i = \mathbb R$ denotes the set of actions of player $i$ and $f_i(x)$ represents the payoff function of player $i$. Let $x_i \in X_i$ denote the action of player $i$ and let $  x := [x_1 \cdots x_N]^T$.  
\end{definition}

\begin{definition}
	(\emph{Nash equilibrium}): An action profile $  x^* = (x^*_i,   x_{-i} ^*) \in X$ is a NE given that 
	\begin{align}
		f_i (x^* _i,   x_{-i} ^*) \le    f_i (x_i,   x_{-i} ^*), \,\, i \in \mathcal N, x_i \in X_i,
	\end{align}
	in which $  x_{-i} : = [x_1\,\, x_2 \cdots x_{i-1}\,\, x_{i+1} \cdots x_N   ] ^T\in \mathbb R ^{N-1}$. 
\end{definition}

By defining 
\begin{align}
	P(  x): = \left[ \frac{\partial f_1 (  x)}{\partial x_1} \cdots \frac{\partial f_N (  x)}{\partial x_N}        \right]^T,
\end{align}
we make the following assumptions:
\begin{assumption}\label{ass 1}
	\cite{9139319} For $i \in \mathcal N$, $f_i( x)$ is a $\mathcal C ^2$ function and $ \frac{\partial f_i (x)}{\partial x_i}$ is globally Lipschitz with constant $l_i$.
	\qedp
\end{assumption}
\begin{assumption}\label{ass 2}
	\cite{9878291} For $  x,  z \in \mathbb R ^ N$, there exists a positive $\mu$ such that 
	$
	(x -  z) ^T (P( x)-P(  z)) \ge \mu \|  x  -  z \|^2. 
	$
	\qedp
\end{assumption}

Assumption \ref{ass 2} is commonly required for ensuring the game to have a unique NE \cite{9878291, chen2022distributed}. We refer the readers to \cite{ye2017distributed, facchinei2003finite} for details.

\subsection{System description}

The players $i=1,2, \cdots, N$ are expressed as linear discrete-time systems given by
\begin{align}\label{system}
	x_i((k+1) \Delta ) =  x_i(k  \Delta  )+ u_i(k \Delta ), k\in \mathbb{Z}_{\ge 0}
\end{align}
where $\Delta$ is the sampling interval and $x_i(k \Delta)\in \mathbb{R}$ denotes the action of player $i$ at time $k\Delta$ with $x_i(0)$ as the initial condition. 
Let $u_i(k) \in \mathbb{R}$ denote the control input, which will be given later. 

In this paper, the communication channel among the players is bandwidth limited and subject to DoS attacks. 
Transmission attempts of players $i=1, 2, \cdots, N$ take place periodically at time $k\Delta$ with $k\in \mathbb{Z}_{\ge 0}$. 
Players $i$ can only exchange information with their neighbor players $j \in \mathcal N_i$. Due to the constraints of network bandwidth, the signals are encoded with a limited number of bits. In the presence of DoS, some of the transmission attempts can fail.



\subsection{Time-constrained DoS}
We refer to DoS as the event for which the encoded signals cannot be received by the decoders when it is active. In our paper, DoS attacks influence the transmissions between the encoders and decoders of all the players simultaneously and hence can cause uniform packet delivering failures across the players.
 We consider a general DoS model
that describes the attacker's action by the frequency of DoS attacks and their duration. Let 
$\{h_q\}_{q \in \mathbb Z_0}$ with $h_0 \geq 0$ denote the sequence 
of DoS \emph{off/on} transitions, that is,
the time instants at which DoS exhibits 
a transition from zero (transmissions are successful) to one 
(transmissions are not successful).
Hence,
$
H_q :=\{h_q\} \bigcup \, [h_q,h_q+\tau_q)  
$
represents the $q$-th DoS time-interval, of a length $\tau_q \in \mathbb R_{\geq 0}$,
over which the network is in DoS status. If $\tau_q=0$, then
$H_q$ takes the form of a single pulse at $h_q$.  
Given $\tau,t \in \mathbb R_{\geq0}$ with $t\geq\tau$, 
let $n(\tau,t)$
denote the number of DoS \emph{off/on} transitions
over $[\tau,t]$, and let 
$
\Xi(\tau,t) := \bigcup_{q \in \mathbb Z_0} H_q  \, \bigcap  \, [\tau,t] 
$
be the subset of $[\tau,t]$ where the network is in DoS status. 

\begin{assumption}\label{ass 3}
	\cite{de2015input}	(\emph{DoS frequency}). 
	There exist constants 
	$\eta \in \mathbb R_{\geq 0}$ as a chattering bound and 
	$\tau_D \in \mathbb R_{> 0}$ specifying the average dwell-time between consecutive DoS off/on transitions in $[\tau, t]$ such that
	$
	n(\tau,t)  \, \leq \,  \eta + (t-\tau)/\tau_D
	$
	for all  $\tau,t \in \mathbb R_{\geq 0}$ with $t\geq\tau$.
	\qedp
\end{assumption}

\begin{assumption} \label{ass 4}
	\cite{de2015input}	(\emph{DoS duration}). 
	There exist constants $\kappa \in \mathbb R_{\geq 0}$ as a chattering bound and $T  \in \mathbb R_{>1}$ characterizing the fraction of the duration of DoS ($\Xi(\tau, t)$) over the total time $t-\tau$ such that
	$
	|\Xi(\tau,t)|  \, \leq \,  \kappa + (t-\tau)/T
	$
	for all  $\tau,t \in \mathbb R_{\geq 0}$ with $t\geq\tau$. 
	\qedp
\end{assumption}

For more discussions regarding Assumptions \ref{ass 3} and \ref{ass 4}, we refer the readers to \cite{feng2017resilient} and \cite{de2015input}. Note that Assumptions \ref{ass 3} and \ref{ass 4} were proposed in 	\cite{de2015input} investigating centralized systems. Later, they have been widely adopted in the works of multi-agent systems, e.g., \cite{feng2019secure,cetinkaya2020randomized}.

For the ease of notation, we let $\{s_r\}_{r\in \mathbb Z_{\ge 0}} \subseteq \{k \Delta\}$ represent the instants of successful transmissions, i.e., $s_r \notin H_q$. Note that $s_0$ is the instant when the first successful transmission occurs, but it is not necessarily the instant of the first transmission attempt if DoS is present at the initial time. Therefore, we let $s_{-1}$ denote the time instant of the first transmission attempt, namely $s_{-1}=0$. 

We assume that the transmissions are free of delay. If DoS is absent, then the decoders can receive the encoded signals immediately at scheduled time $k\Delta$.
If players do not receive any information from the others at scheduled time $k \Delta$, this implies that DoS is present at $k \Delta$. By this, players are able to locally check the status of DoS attacks. At last, the communication channel is assumed to be free of noise, errors and random dropouts. If random dropouts and DoS are co-present, one may need to follow a stochastic type of analysis. For more information, we refer the readers to \cite{cetinkaya2016networked}.

Let $T_S(0,k\Delta)$ denote the number of successful transmissions between 0 and $k\Delta$. The lemma below presents a lower bound of $T_S(0,k\Delta)$.

\begin{lemma} \label{Lemma T} 
 \cite{feng2023tcns} Consider the DoS attacks characterized by Assumptions \ref{ass 3} and \ref{ass 4} and the network sampling period $\Delta$. If 
	$1/T+ \Delta /\tau_D <1$, then 
	$
	T_S(0, k\Delta))
	\ge  \left(1- 1/T - \Delta/\tau_D \right) k   - \frac{\kappa+\eta\Delta}{\Delta}.  
	$ \qedp
\end{lemma}

\subsection{Quantizer}
The limitation of bandwidth implies that transmitted signals are subject to quantization. Let $b \in \mathbb{R}$ be the original scalar signal before quantization and $q_R(\cdot)$ be the quantization function for a scalar
input value as
\begin{align}\setlength{\arraycolsep}{3pt}  \label{quantizer}
	q_R (b) = 
	\left\{
	\begin{array}{lll}
		0 & -\frac{1}{2} < b < \frac{1}{2} & \\
		z  & \frac{2z-1}{2} \le b <  \frac{2z+1 }{2} \\
		R & b \ge    \frac{2R+1}{2}&         \\
		-q_R (-b) & b \le -\frac{1}{2} & 
	\end{array}
	\right.
\end{align}
where $R\in \mathbb{Z}_{>0}$ is to be designed and $z =1, 2, \cdots, R$. If the quantizer is unsaturated such that $b \le (2R+1)/2 $, then the error induced by quantization satisfies 
$
	|b - q_R(b)| \le 1/2$.
Moreover, we define the vector version of the quantization function as $Q_R(y) = [\,q_R(y_1)\,\,q_R(y_2)\,\, \cdots \,\, q_R(y_p) \, ]^T \in \mathbb R ^p$ with $y=[y_1 \cdots y_p]^T  \in \mathbb R ^p$. 


\subsection{Control objectives}
This paper has two control objectives: 
\begin{itemize}
	\item[1.] Design a quantized distributed NE seeking controller and find a $R$ (or equivalently $(2R+1)/2$) such that the quantizer (\ref{quantizer}) is not saturated at all $k$, i.e., $|b| \le (2R+1)/2$. 
	\item[2.] The actions of the players converge to the NE under DoS attacks, i.e., $ x(k\Delta) \to   x^*$ as $k \to \infty$.
\end{itemize}
Note that when quantizer saturation occurs, a change of $b$ in (\ref{quantizer}) will not be reflected by $q_R(b)$. From an intuitive viewpoint, under saturation, the feedback mechanism fails. Thus, Control objective 1 is important.

\section{quantized distributed NE seeking controller design}


\subsection{Control architecture}
To reduce notation burden, we let $k$ represent $k \Delta$ in the remainder of the paper, e.g., let $x_i(k )$ represent $x_i(k \Delta )$.  

Recall that $k\notin H_q$ means DoS is not active at $k$, and $k\in H_q$ vice versa. We propose the control input to player $i$ in (\ref{system}) as 
\begin{align}\label{u}
	u_i (k) = 
	\left\{
	\begin{array}{ll}
		-\delta \frac{\partial f_i}{\partial x_i}(x_i(k),   y_{-i}(k))& k \notin H_q\\
		0  & k \in H_q
	\end{array}
	\right.
\end{align}
where $\delta >0$ is a design parameter given later, $  y_{-i}(k) = [y_{i1}(k)\,\, y_{i2}(k)\,\, \cdots\,\, y_{i,i-1}(k)\,\, y_{i,i+1}(k) \cdots  y_{iN}(k)]^T  \in \mathbb R^{N-1}$ denotes the estimation of the actions of all the other players in the controller of player $i$ and $\frac{\partial f_i}{\partial x_i}(x_i(k),y_{-i}(k))=\frac{\partial f_i(x_i(k),x_{-i}(k))}{\partial x_i}|_{x_{-i}(k)=y_{-i}(k)}$. For instance, $y_{ij}(k)$ denotes the estimation of $x_j(k)$ by player $i$. Specifically, for $y_{ij}(k)$ with $i, j \in \mathcal N$, we propose 
\begin{subequations}\label{yij}
	\begin{align}
		&y_{ij}(k+1) = y_{ij}(k) -  \sigma (k) \times \nonumber\\
		& \quad  \left(\sum_{l=1}^{N}a_{il}(\hat y_{ij}(k)- \hat y_{lj}(k)) + a_{ij}(\hat y_{ij}(k)- \hat x_j(k))\right)  \\
		&\sigma(k)= \left\{
		\begin{array}{ll}
			h &    k \notin H_q\\
			0  & k \in H_q
		\end{array}
		\right.
	\end{align} 
\end{subequations}
in which $h$ is a design parameter specified later in Lemma \ref{lemma H} and we let $y_{ij}(0)=0$ for the ease of subsequent computation. 
To understand (\ref{yij}) intuitively, one can regard (\ref{yij}) as a quantized leader-follower consensus protocol in the presence of DoS attacks, in which $j$ is the leader and all the players other than $j$ are the followers. Namely, $y_{ij}$ should track $\hat x_j$ by (\ref{yij}), which requires that player $j$ must generate $\hat x_j$ and send it to the neighbors of $j$. Overall, if $y_{ij}\to\hat x_j$ by the consensus protocol (\ref{yij}) and $\hat x_j \to x_j$ by a proper quantization mechanism design, then one realizes $y_{ij}\to x_j$, i.e., player $i$ estimates the action of player $j$ successfully.

In our paper, the encoder and decoder for each state have the same structure (i.e., (\ref{hat yij})-(\ref{eq h})). 
In the following, we explain the decoding process, i.e., how to obtain the values of $\hat y_{ij}$, $\hat y_{lj}$ and $\hat x_j$ in (\ref{yij}). The encoding process is similar to the decoding process and hence will be omitted due to space limitation.

The dynamics of $\hat y_{ij}(k)$ in (\ref{yij}) follows
\begin{subequations}\label{hat yij}
	\begin{align}
		&\hat y_{ij}(k+1) \!\!=\!\!
		\left\{\!\!\!
		\begin{array}{ll}
			\hat y_{ij}(k) + \theta(k) \hat Q_{ij}(k+1) & \! k +1  \!\notin\! H_q\\
			\hat y_{ij}(k) &\!   k+1 \!\in\! H_q\\
		\end{array}
		\right.\\
		&\hat Q_{ij}(k+1) = q_R \left(\frac{y_{ij}(k+1) -    \hat y_{ij}(k)}{\theta(k)}  \right)
	\end{align}
\end{subequations}
in which $\theta(k)$ is the scaling parameter specified later in (\ref{eq h}). 
The state $\hat y_{lj}(k)$ in (\ref{yij}) is computed 
in a very similar way as in (\ref{hat yij}). The dynamics of $\hat x_j(k)$ in (\ref{yij}) follows
\begin{subequations}\label{hat xj}
	\begin{align}
		&\hat x_{j}(k+1) \!\!=\!\!
		\left\{\!\!
		\begin{array}{ll}
			\hat x_{j}(k) +\theta(k) \hat Q_{j}(k+1) &\!\!  k +1 \notin H_q\\
			\hat x_{j}(k) &\!\!  k+1 \in H_q\\
		\end{array}
		\right.\\
		&\hat Q_{j}(k+1) = q_R \left(\frac{x_{j}(k+1) -    \hat x_{j}(k)}{\theta(k)}  \right).
	\end{align}
\end{subequations}
For the initial conditions of (\ref{hat yij}) and (\ref{hat xj}), we let $\hat y_{ij} (0)= 0$ and $\hat x_j (0)= 0$, respectively, for the ease of later computation.

In order to ensure that the state is always within the bounded quantization range, i.e., free of quantizer saturation, each player needs to adjust the value of the scaling parameter $\theta (k)  \in \mathbb{R}_{>0} $ dynamically in (\ref{hat yij}) and (\ref{hat xj}), respectively. Specifically, we need to ensure that $|(y_{ij}(k+1)- \hat y _{ij}(k)) / \theta(k)|$ and $|(x_j(k+1)- \hat x _j (k)) / \theta(k)|$ in (\ref{hat yij}) and (\ref{hat xj}), respectively, are always no larger than $(2R + 1) /2$ in (\ref{quantizer}). Otherwise, quantizer saturation occurs. 
The scaling parameter $\theta(k)$ is updated by the distributed laws:
\begin{align}\label{eq h}
	\theta(k) &= 
	\left\{
	\begin{array}{ll}
		\gamma_1 \theta(k-1) & \,\,\,\, \text{if $k  \notin H_q $} \\
		\gamma_2 \theta(k-1) & \,\,\,\, \text{if $k  \in H_q $  }
	\end{array}
	\right.  \,\,  k=1, 2\cdots.
\end{align}
with $\theta(0) = \theta_0 \in \mathbb{R}_{>0}$. In this paper, we select the zooming-in factor $\gamma_1 <1$ and the ``zooming-out factor" $\gamma_2 =1$. Since $\gamma_2=1$, ``zooming-out" also implies ``holding". The reasons of such $\gamma_1<1$ for zooming-in and $\gamma_2=1$ for holding will be specified later in Remark \ref{remark 3}. Please note that acknowledgments are not necessary because each player is able to passively know the status of DoS at $k$ by whether receiving the neighbors' transmissions. Thus, $\theta(k)$ can be updated locally. Since each player has the identical initial condition $\theta_0$, they can also locally check $k \in H_q$ or $k \notin H_q$, one can verify that $\theta(k)$ is synchronized in all the players.

In (\ref{eq h}), the update of $\theta(k)$ has two modes. The first mode is for NE seeking and the second is particularly for mitigating the influence of DoS attacks. Under DoS attacks, $y_{ij}(k+1) -    \hat y_{ij}(k)$ and $x_j(k+1)  - \hat x_j (k)$ in (\ref{hat yij}) and (\ref{hat xj}) can diverge \cite{feng2023tcns}. If $\theta(k)$ always zooms in by only utilizing $\gamma_1$ as in \cite{chen2022distributed}, then as $\theta(k)$ approaches zero, $|(y_{ij}(k+1)- \hat y _{ij}(k)) / \theta(k)|$ and $|(x_j(k+1)- \hat x _j (k)) / \theta(k)|$ in $q_R(\cdot)$ can approach infinity. Eventually, quantizer saturation will occur, e.g., $|(y_{ij}(k+1)- \hat y _{ij}(k)) / \theta(k)|>(2R+1)/2$. 
Quantizer saturation will also occur in general if the players transmit quantized $y_{ij}(k+1)/\theta(k)$ to other players instead of following the mechanism in (\ref{hat yij}), even without DoS. This is because the growing $y_{ij}(k+1)/\theta(k)$ (due to $\theta(k)\to 0$) will exceed the range of any finite quantizer $q_R(\cdot)$. The same problem also exists for (\ref{hat xj}).

In \cite{ye2017distributed}, the exact values of $y_{ij}(k)$, $y_{lj}(k)$ and $x_j(k)$ can be received by players since the network bandwidth is implicitly assumed to be unlimited. However, in case bandwidth is limited, the exact values of $y_{ij}(k)$, $y_{lj}(k)$ and $x_j(k)$ are not available. Instead, players need to estimate them from quantized data by (\ref{hat yij}) and (\ref{hat xj}). In \cite{9878291}, the range of quantizer is not provided, which implicitly assumes that quantizer range is sufficiently large without saturation problem. In our paper, as one of the control objectives, we will specify a lower bound of quantizer range preventing saturation.

\begin{algorithm}[t]
 
		\caption{Computation tasks and updates of player $i$} \label{alg:cap}
	\begin{algorithmic}[1] 
		\Require   $\theta(0)=\theta_0$, $\delta$ and $\gamma_1$ in Lemma \ref{lemma H}, $\gamma_2=1$, $h$ in Lemma \ref{lemma: h}, $\hat y_{ij}(0)=0$, $\hat x_j(0)=0$, $y_{ij}(0)=0$, $x_i(0)=0$ and $q_R(\cdot)$ 
		\While{$k=0, 1, 2\cdots$, }
		\If{$k\notin H_q$} \textbf{compute}
		\State 	 $u_i(k)$ by the 1st eq. in (\ref{u}),\;
		\State 	 $\theta(k)$ by the 1st eq. in (\ref{eq h}),*\;
		\State 		  $\hat  y_{lj}(k)$  the 1st eq. in (\ref{hat yij}a) when receive $\hat Q_{lj}(k)$ ($\forall  l\in \mathcal N_i$),*\;
		\State 		 $\hat Q_{ij}(k)$ and $\hat y_{ij}(k)$ by the 1st eq. in (\ref{hat yij}a),*\;
		\State 		 $\hat  x_{j}(k)$ by the 1st eq. in (\ref{hat xj}a) when receive $\hat Q_{j}(k)$ (if $j\! \in\! \mathcal N_i$),*\;
		\State 		Select $\sigma(k)=h$ by the 1st eq. in (\ref{yij}b),\;
		\ElsIf{$k\in H_q$}
		\State	select $u_i(k)=0$ by the 2nd eq. in (\ref{u}),\;	
		\State		 compute $\theta(k)$ by the 2nd eq. in (\ref{eq h}),*   \;
		\State			 compute $\hat y_{lj}(k)$  ($\forall l\in \mathcal N_i$)  by the 2nd eq. in (\ref{hat yij}a),*   \;
		\State			 compute $\hat x_{j}(k)$ (if $j \in \mathcal N_i$) by the 2nd eq. in (\ref{hat xj}a),*\; 
		\State			select $\sigma(k)=0$ by the 2nd eq. in (\ref{yij}b). \;
		\EndIf
		\State 		compute $y_{ij}(k+1)$ by (\ref{yij}) ($\forall j\in \mathcal N/i$),  \; 
		 \State 	compute $x_i(k+1)$ by (\ref{system}).\; 
		\EndWhile 
			 	\end{algorithmic}
		 	\Comment{Lines with ``$*$" are not needed at $k=0$. 
		 	 \quad	\quad\quad\quad	\quad\quad\quad}
\end{algorithm}


By (\ref{u})--(\ref{eq h}), one can see that the NE seeking controller has two modes depending on the status of DoS attacks. We propose such a switched-type controller for mitigating the influence of DoS attacks. The laws for quantized NE seeking in (\ref{yij})--(\ref{eq h}) are partially inspired by the approaches in quantized consensus \cite{li2010distributed, you2011network, feng2023tcns}. However, as will be shown later, due to the presence of nonlinear function in (\ref{u}), the detailed technical analysis is different from quantized consensus. 


\subsection{Quantized system dynamics}

We define a few vectors to facilitate the subsequent analysis.\\
Quantization error of $x(k)$, i.e., $e  ^ x(k)$:
\begin{subequations}\label{11}
	\begin{align}
		& e_i ^ x (k) 	: = x_i (k) - \hat x_i (k)  \in \mathbb R  \\
		&e  ^ x (k) : = [e_1 ^ x (k) \cdots e_N ^ x (k)]  \in \mathbb R ^N.
	\end{align}
\end{subequations}
Compact form of $y(k)$:
\begin{subequations}\label{14}
	\begin{align}
		&y_i (k) : = [y_{i1}(k) \,\, \cdots\,\, y_{iN}(k)]^T \in \mathbb R ^{N} \\
		&y (k) : = [y_{1} ^T(k) \,\, \cdots\,\, y_{N} ^T(k) ]^T \in \mathbb R ^{N^2}. 
	\end{align}
\end{subequations}
Quantization error of $y(k)$, i.e, $e  ^y (k)$:
\begin{subequations}
	\begin{align}
		&e_{ij} ^y (k): = y _{ij} (k) - \hat y _{ij} (k) \in \mathbb R  \\
		&e_i ^y (k) : = [e_{i1} ^y (k) \cdots e_{iN} ^y (k)] \in \mathbb R ^ N  \\
		&e  ^y (k): =[e_1 ^y (k) ^ T \cdots e_N ^y (k) ^ T ]  \in \mathbb R ^{N^2}.   
	\end{align}
\end{subequations}
Estimation error of $x(k)$, i.e., $\bar y(k)$:
\begin{subequations}\label{bar y}
	\begin{align}
		&x(k) := [x_{1} (k) \,\, \cdots\,\, x_{N}(k)]^T \in \mathbb R ^{N} \\
		&\bar y(k) : = y(k) - \mathbf 1 _N \otimes x(k) \in \mathbb R ^{N^2}.
	\end{align}
\end{subequations}
The variable $\bar y$ can be interpreted as follows. In the absence of quantization (e.g., in \cite{ye2017distributed}), $\bar y$ denotes the discrepancy between the ``real state" $x$ and its estimation $y$ by players. In our paper, it has a similar meaning whereas in the presence of signal quantization. It will be shown later that $\bar y(k) \to 0$ as $k \to \infty$.

Based on the status of DoS attacks, the dynamics of the players can be formulated by four cases: 
\begin{itemize}
	\item Case 1) $k \notin H_q$ and $k+1 \notin H_q$: consecutive successful transmissions
	\item Case 2) $k \notin H_q$ and $k+1 \in H_q$: DoS from absence to presence
	\item Case 3) $k \in H_q$ and $k+1 \in H_q$: consecutive unsuccessful transmissions
	\item Case 4) $k \in H_q$ and $k+1 \notin H_q$: DoS from presence to absence. 
\end{itemize}
System dynamics corresponding to the four cases above are as follows:

Case 1) To obtain (\ref{15}), it is important to note the following dynamics for Case 1): $u_i(k)=-\delta \frac{\partial f_i}{\partial x_i}(x_i(k),   y_{-i}(k))$ by (\ref{u}) and hence $x_i(k+1)= x_i(k)-\delta \frac{\partial f_i}{\partial x_i}(x_i(k),   y_{-i}(k))$, $\sigma(k)=h$ by (\ref{yij}b) and hence $y_{ij}(k+1) = y_{ij}(k) -  \sigma (k)  (\sum_{l=1}^{N}a_{il}(\hat y_{ij}(k)- \hat y_{lj}(k)) + a_{ij}(\hat y_{ij}(k)- \hat x_j(k)) )$ by (\ref{yij}a), $\hat{y}_{ij}(k+1)=\hat y_{ij}(k) + \theta(k) \hat Q_{ij}(k+1)$ by (\ref{hat yij}a), $\hat x_j(k+1)= 	\hat x_{j}(k) + \theta(k) \hat Q_{j}(k+1) $ by (\ref{hat xj}a). Then, one can obtain the iterative equations of $\bar y(k)$, $x(k)$, $e^x(k)$ and $e^y(k)$ by the dynamics above and the definitions in (\ref{11})--(\ref{bar y}): 
\begin{subequations}\label{15}
	\begin{align}
		&\!\! \bar y(k+1) = H\bar y(k) +  \delta  \mathbf 1 _N  \otimes P(\eta(k))  \nonumber  \\ 
		&\!\! \quad  \quad \quad  \quad+   hS e ^ y (k)	- h A_0 (\mathbf 1 _N \otimes e^ x(k))  \\
		&\!\! x(k+1) - x^* = x(k) - x^* - \delta P(\eta(k)) \\
		&\!\!\!e ^ x (k+1) =  e^x (k) -  \delta P (\eta(k))  \nonumber\\
		& - \theta(k) Q _R \left( \frac{e^x (k) -  \delta P (\eta(k))}{\theta(k) }\right) \\
		&\!\! e^ y (k+1) = G e^y (k) - h A_0 (\mathbf 1 _ N \otimes e^x (k)) - h S \bar y(k) \nonumber\\
		&   - \theta(k)Q_R \left(\! \frac{G e^y (k) \!-\! h A_0 (\mathbf 1 _ N \otimes e^x (k)) \!-\! hS \bar y(k)  }{\theta(k)}\!\right)
	\end{align}
\end{subequations}
in which
\begin{align*}
	&  H \!: =\! I _{N^2} \!-\! h ( \mathcal L \otimes I_N \!+\! A_0) \label{17}, \,\,\,\,\,\,\,\,  G \!: =\! I _ {N^2} \!+\! h (\mathcal L \otimes I_N \!+\! A_0)\\
	&  S \!: =\! \mathcal L \!\otimes\! I_N \!+\! A_0,\,\,\,\,\,\,\,
	A_0\!\!:=\! \text{diag}[a_{11}\!\cdots\! a_{1N}\,\,a_{21}\!\cdots\! a_{2N}\!\cdots\! a_{NN}]\\
	&  P(\eta(k)) \!\!:= \!\!   \left[ \frac{\partial f_1}{\partial x_1}(x_1(k),   y_{-1}(k)) \!\cdots\! \frac{\partial f_N}{\partial x_N}(x_N(k),   y_{-N}(k))     \right]^T\!\!.  
\end{align*}

In the following lemma, we present the selection of design parameter $h$ in (\ref{yij}b).

\begin{lemma}\label{lemma: h}
	\cite{9139319} For the matrix $H$ in (\ref{15}a), if
	\begin{align}
		h < \min _{i, j\in \mathcal N} \frac{1}{\sum_{l=1}^{N} a_{il} + a_{ij}}
	\end{align}
	then all the eigenvalues of $H$ are within the unit circle. \qedp
\end{lemma}

Since $\mathcal L$ is symmetric and $A_0$ is diagonal, it is simple to verify that $H$ is a Hermitian matrix. Then we have $\|H\| = \rho(H)<$ 1. This property will facilitate the analysis in Appendix-C.

Case 2)
	One should note that: $u_i(k)$, $x_i(k+1)$, $y_{ij}(k+1)$ follow the same dynamics as in Case 1) because of $k \notin H_q$. However, one has $\hat y_{ij}(k+1)= \hat y_{ij}(k)$ by (\ref{hat yij}a) and $\hat x_j(k+1)= \hat x_j(k)$ by (\ref{hat xj}a) because of $k+1\in H_q$. Then, one has
\begin{subequations}\label{Case 2}
	\begin{align}
		&\bar y(k+1) = H\bar y(k) +  \delta  \mathbf 1 _N  \otimes   P(\eta(k))  \nonumber  \\ 
		& \quad  \quad  \quad  \quad  \quad +   hS e ^ y (k)	- h A_0 (\mathbf 1 _N \otimes e^ x(k))  \\
		& x(k+1) - x^* = x(k) - x^* - \delta P(\eta(k))\\
		&e ^ x (k+1) =  e^x (k) -  \delta P (\eta(k))   \\
		&e^ y (k+1) = G e^y (k) - h A_0 (\mathbf 1 _ N \otimes e^x (k)) - h S \bar y(k). 
	\end{align}
\end{subequations}

Case 3) To obtain (\ref{Case 3}), one has $u_i(k)=0$ by (\ref{u}) and hence $x_i(k+1)=x_i(k)$, $\sigma(k)=0$ and hence $y_{ij}(k+1)= y_{ij}(k)$ by (\ref{yij}) because of $k \in H_q$; $\hat y_{ij}(k+1)$ and $\hat x_j(k+1)$ follow the same dynamics as in Case 2) because of $k+1\in H_q$. Then, we obtain
\begin{subequations}\label{Case 3}
	\begin{align}
		&\bar y(k+1) =  \bar y(k) \\
		& x(k+1) - x^* = x(k) - x^*  \\
		&e ^ x (k+1) =  e^x (k)   \\
		&e^ y (k+1) =   e^y (k).  
	\end{align}
\end{subequations}

Case 4)  To obtain (\ref{Case 4}), first note that $x_i(k+1)=x_i(k)$ and $y_{ij}(k+1)= y_{ij}(k)$ due to $k\in H_q$ (see Case 3). Moreover, one should also note that $\hat{y}_{ij}(k+1)=\hat y_{ij}(k) + \theta(k) \hat Q_{ij}(k+1)$ by (\ref{hat yij}a), $\hat x_j(k+1)= 	\hat x_{j}(k) + \theta(k) \hat Q_{j}(k+1) $ by (\ref{hat xj}a) due to $k+1\notin H_q$. Then, one has
\begin{subequations}\label{Case 4}
	\begin{align}
		&\bar y(k+1) =  \bar y(k) \\
		& x(k+1) - x^* = x(k) - x^*  \\
		&e ^ x (k+1) =  e^x (k)  - \theta(k) Q_R \left( \frac{e^x(k)}{\theta(k)}\right)  \\
		&e^ y (k+1) =   e^y (k)   -  \theta(k) Q_R \left( \frac{e^y(k)}{\theta(k)}\right).
	\end{align}
\end{subequations}

It is clear that if the network is free from DoS attacks, Case 1) is sufficient for analyzing the quantized distributed NE seeking problem. The presence of DoS attacks complicates the NE seeking problem by introducing Cases 2)-4). Consequently, the analysis becomes more challenging due to  Cases 2)-4) as shown in the proofs of Lemma \ref{lemma H} and Theorem \ref{theorem} later. Compared with quantized consensus of linear systems, the technical challenge is raised by the involvement of the nonlinear function $P(\eta(k))$.

\section{Main results}

In this section, we will show that our control scheme in Section III can realize the control objectives in Section II-D. 


\subsection{Analysis of zooming-in and holding quantization }
We follow the approach aligned with that for quantized consensus under data rate constraints \cite{li2010distributed, you2011network, feng2023tcns} and define
\begin{subequations}\label{transformation}
	\begin{align}
		&\beta (k) : = \frac{\bar y(k)}{\theta(k)} ,\quad \chi (k) : = \frac{x(k) - x^*}{\theta(k)}    \\
		&\xi ^ y (k) : = \frac{e^ y (k)}{\theta(k)}, \quad
		\xi ^ x (k) : = \frac{e^ x (k)}{\theta(k)}.    
	\end{align} 
\end{subequations}
Recall that $s_r$ denotes the instant of successful transmissions. Then one has $\|\xi^x(s_r)\|_\infty \le 1/2\gamma_1$ and $\|\xi^y(s_r)\|_\infty \le 1/2\gamma_1$ if the quantizer $Q_R(\cdot)$ does not saturate (see Appendix-A).

For quantized consensus without DoS attacks \cite{li2010distributed, you2011network}, to obtain a quantization range preventing saturation and achieve consensus, it is necessary to derive an upper bound of consensus error (discrepancy between states) scaled by the scaling parameter, by utilizing the iteration between $k$ and $k+1$.  In our paper, we would intend to obtain an upper bound of $x(k)-x^*$ scaled by the scaling parameter $\theta(k)$, namely $\chi(k)$. However, we cannot obtain the bound by using the iteration of $\chi(\cdot)$ between $k$ and $k+1$ due to the presence of DoS and the consequent Cases 1)-4). Therefore, we attempt to derive an upper bound of $\chi(\cdot)$ at successful transmission instants by exploiting the iteration between $\chi(s_r)$ and $\chi(s_{r+1})$. Because $\chi(k)$ is coupled with $\beta(k)$, one needs to consider their dynamics simultaneously. 
Then, we let
\begin{align}\label{pi}
\|\pi(k)\| =	\left\|	\begin{bmatrix} 
	\|   \beta (k)  \| \\
	\|\chi (k)  \|
\end{bmatrix}\right\|\,\, \text{with}\,\,
\pi(k):=		\begin{bmatrix} 
	\|   \beta (k)  \| \\
	\|\chi (k)  \|
\end{bmatrix}.
\end{align}

\begin{assumption}\label{Assumption 5}
We assume that there exist $C_{x_0}$ and $C_{x^*}$ such that $ C_{x_0} \ge |x_i(0)| $ and $C_{x^*}\ge |x^* _i |$ for $i \in \mathcal N$, respectively. \qedp
\end{assumption}

Note that a similar assumption on the initial condition is also required in the literature of quantized consensus \cite{you2011network, li2010distributed}. For quantized NE seeking problems, one also needs the assumption $C_{x^*} \ge |x^* _i |$. They both are required for calculating an upper bound of players' actions and preventing quantizer saturation at the initial time.  
We refer the readers to Remark 1 in \cite{chen2022distributed} for more detailed reasons.

The following lemma specifies the selection of the zooming-in parameter $\gamma_1$ in (\ref{eq h}) and an upper bound of $\|\pi(s_r)\|$. 
Its proof is provided in Appendix-B. 

\begin{lemma}\label{lemma H}
Suppose that Assumption \ref{Assumption 5} holds. Select $\gamma_1 \in (\rho(\bar H), 1)$ with
	\begin{align}
		\bar H:=
		\begin{bmatrix}	
			\|H\| + \delta    l \sqrt{N} &  \delta     l N \\
			\delta    l N  & \sqrt{1-2\delta   \mu + (\delta   l)^2}
		\end{bmatrix}
	\end{align}
	where $l:=\max\{l_i\}_{i \in \mathcal N}$ (given in Assumption \ref{ass 1}).
	If $\|\xi ^y (s_g)\|_\infty \le 1/2\gamma_1$ and $\|\xi ^x (s_g)\| _\infty \le 1/2\gamma_1$ for $g=-1, 0, \cdots, r-1$, then 
	\begin{align}\label{value C}
		\|\pi(s_r)\| & \le  \max\left\{\frac{\bar C_\gamma}{\theta_0}\sqrt{N^2 C_{x_0} ^2 + N (C_{x^*}+ C_{x_0})^2}, \right.\nonumber\\
		&\quad\quad\quad\quad\left.\frac{ C_{\gamma}hN(\|S\|+\|A_0\|)}{2\gamma_1^2(1-\gamma)} \right\}=:C 
	\end{align}
	where $\rho(\bar H)<1$, $\bar C_\gamma : = \max\{\gamma C_{\gamma} ,1\}$. Here, $C_{\gamma} \in \mathbb R _{\ge 1}$ and $\rho(\bar H)/ \gamma_1 \le \gamma<1$ satisfy $\| (\bar H  /\gamma_1)^k \|\le C_{\gamma} \gamma ^ k$. The existence of $\delta$ realizing $\rho(\bar H)<1$ is presented in Appendix-C. \qedp
\end{lemma}

\begin{remark}\label{remark 2}
1) Interpreting Lemma \ref{lemma H} from an intuitive aspect: First, recall that $\|\xi^x(s_r)\|_\infty \le 1/2\gamma_1$ and $\|\xi^y(s_r)\|_\infty \le 1/2\gamma_1$ if $Q_R(\cdot)$ does not saturate in view of (\ref{21}) and (\ref{22}). Then, if the quantizer is not saturated at the instants of successful transmissions before $s_r$ (i.e., $\|\xi ^y (s_g)\|_\infty \le 1/2\gamma_1$ and $\|\xi ^x (s_g)\| _\infty \le 1/2\gamma_1$ for $g=-1, 0, \cdots, r-1$ in Lemma \ref{lemma H}), then at $s_r$, $\|\pi(\cdot)\|$ will be upper bounded by $C$.

2) The purposes of Lemma \ref{lemma H} are two-fold: Specify the zooming-in parameter $\gamma_1$ and provide an upper-bound on $\|\pi(s_r)\|$. Note that $0<\gamma_1<1$ is necessary for the success of NE seeking even without DoS attacks. The result $\|\pi(s_r)\|<C$ is also necessary since we will use it to show that $\|\pi(\cdot)\|$ is upper bounded for all $k$. Then one can obtain the main results: Quantizer unsaturation at all $k$ and NE seeking as $k\to \infty$. 


3) The induction-type proof of Lemma \ref{lemma H} follows a simple idea: Suppose that $0=s_{-1} < s_0$. If $\|\xi^y(s_{-1})\|\le 1/2\gamma_1$ and $\|\xi^x(s_{-1})\|\le 1/2\gamma_1$, then $\|\pi(s_0)\|$ will be upper bounded. Furthermore, if $\|\xi^y(s_g)\|\le 1/2\gamma_1$ and $\|\xi^x(s_g)\|\le 1/2\gamma_1$ ($g=-1, 0$) hold, then $\|\pi(s_1)\|$ will be upper bounded. Extending to the general case, if $\|\xi^y(s_g)\|\le 1/2\gamma_1$ and $\|\xi^x(s_g)\|\le 1/2\gamma_1$ hold ($g=-1, 0, \cdots, r-1$), then $\|\pi(s_r)\|$ will be upper bounded, as shown in Lemma \ref{lemma H}. The case of $0=s_{-1} = s_0$ can be conducted in a similar way.  


4) Compared with quantized consensus, computing the zooming-in factor $\gamma_1$ is one of the major technical challenges due to the presence of the nonlinear function $P(\eta(k))$ (caused by the nonlinear payoff function $f_i(x)$). In quantized consensus problems, the computation of $\gamma_1$ is more straightforward by exploiting the spectral radius of the dynamic matrix of consensus error \cite{li2010distributed, you2011network}. By contrast, such a  method is not applicable in our paper because the ``NE convergence error" $\|x(k+1)-x^*\|$ is coupled with $\|\bar y(k+1) \|$, and hence the dynamic matrix becomes $\bar H$. To select $\gamma_1$, it is necessary that the spectral radius $\rho(\bar H)<1$. However, the spectral radius of $\bar H$ is not straightforward. Moreover, due to  $\sqrt{1-2\delta   \mu + (\delta   l)^2}$ in $\bar H$, the direct method to characterize $\delta$ for $\rho(\bar H)<1$, commonly used in distributed optimization/NE seeking \cite{xiong2022quantized, meng2023linear}, is not applicable. Therefore, we develop new techniques to characterize $\delta$ for $\rho(\bar H)<1$ indirectly, i.e., by inspecting the derivative of a function (related to $\rho(\bar H)$) along $\delta$ (see Appendix-C).  
\qedp
\end{remark}

\subsection{Quantizer unsaturation and convergence to the NE}

We are ready to present the main results of the paper. 

\begin{theorem}\label{theorem}
	If DoS attacks in Assumptions \ref{ass 3} and \ref{ass 4} satisfy $1/T + \Delta / \tau_D < 1 $, we have the following results:
	
	i) The quantizer (\ref{quantizer}) is not saturated if the quantizers for quantizing $x_i$ and $y_{ij} $ satisfy:
	\begin{subequations}\label{result theorem}
		\begin{align}
			&\text{For $x_i$: }  \frac{2R + 1}{2} \ge  \frac{N}{2\gamma_1} + (1+\sqrt{N})   \delta   l C \label{36} \\
			&\text{For $y_{ij}$: }  \frac{2R + 1}{2} \ge  \frac{N\|G\|}{2\gamma_1} + h \|A_0\|  \frac{N}{2\gamma_1}  + h \|S\| C   \label{37}
		\end{align}
	\end{subequations}
	where $C$ has been provided in Lemma \ref{lemma H}.
	
	ii) The actions of the players converge to the NE $x^*$.
\end{theorem}

\emph{Proof.}
First, we present the basic ideas of the proof. For result i), its proof is conducted by induction: If the quantizer is not saturated at $s_p$ ($p=-1, 0, \cdots, r$, where DoS is present at $k=0$; $p= 0, \cdots, r$, where DoS is absent at $k=0$), then it will not saturate for the transmissions at $k \in (s_r, s_{r+1}]$. This also manifests that the quantizer is not saturated at $s_{r+1}$, which implies that the quantizer will not saturate for $k \in (s_{r+1}, s_{r+2}]$. By induction, one can verify that the quantizer is not saturated at any $k$ in $s_{l}\bigcup_{r \in \mathbb Z_{\ge l} }(s_r, s_{r+1}]$ ($l =-1$ if DoS is present at $k=0$; $l=0$ if DoS is absent at $k=0$), which are essentially all the transmission attempts.
For obtaining result ii), it is sufficient to show that $\|\pi(k)\|$  is upper bounded and the scaling function $\theta(k)\to 0$ by the fact $	\left\| \! 
\begin{bmatrix} 
	\|   \bar y (k)  \| \\
	\| x(k) - x^*  \|
\end{bmatrix}
\! 
\right\|
 =  \theta(k)
\|\pi(k)\| $. In particular, the scaling function $\theta(k)\to 0$ if DoS attacks satisfy  $1/T + \Delta / \tau_D < 1 $.

In the following, Steps 1-4 prove the unsaturation of the quantizer (result i)) and Step 5 presents the success of NE seeking (result ii)) .

\textbf{Step 1.} 
To show the quantizer is free of saturation, it is sufficient to check if $\| y (k+1)  - \hat y (k)\|_\infty /\theta(k)  $ and $\| x (k+1)  - \hat x (k)\|_\infty  /\theta(k)$  in $Q_R (\cdot)$ (see (\ref{hat yij}b) and (\ref{hat xj}b), respectively) are bounded by the maximum quantization range $(2R+1)/2$. 
We conduct an induction-type analysis to show this, i.e., if the quantizers satisfying (\ref{result theorem}) are not saturated at $s_{-1}, s_{0}, \cdots s_{r}$, then they will not saturate at $k \in (s_r, s_{r+1}]$, which essentially implies that the quantizers are not saturated at all $k$.

Recall the definition of $s_r$. Then, for $s_r  + 1$, one needs to investigate $\| y (s_r+1)  - \hat y (s_r)\|_\infty  /\theta(k) $ and $\| x (s_r+1)  - \hat x (s_r)\|_\infty  /\theta(k) $. One can verify that  
\begin{align}\label{step 1 a}
	&\| y (s_r+1)  - \hat y (s_r)\|_\infty   /\theta(s_r) \nonumber\\
	&\le \| y (s_r+1)  - \hat y (s_r)\|    /\theta(s_r)  \nonumber\\
	&\le \|        G e^y (s_r)  - h A_0 (\mathbf 1 _N \otimes e^x (s_r))  - h S \bar y(s_r)   \|  /\theta(s_r)  \nonumber\\
	&\le \|G\| \|\xi ^ y (s_r)\| + h \|A_0\| \sqrt{N} \|\xi ^ x (s_r)\| + h \|S\| \|\beta(s_r)\|    \nonumber\\
	& \le  N\|G\|/2\gamma_1 \!+ \! h \|A_0\|  N/2\gamma_1  \!+\! h \|S\| C    \le  (2R+1)/2
\end{align}
in which 
one has obtained
$
\|\xi ^ y (s_r)\|  \le \frac{N}{2\gamma_1}$,
$ \|\xi ^ x (s_r)\| \le  \frac{\sqrt{N}}{2\gamma_1}$ (after (\ref{compact 2}) in Appendix-A)
and 
$
\|\beta(s_r)\|  \le \|\pi(s_r)\| \le C
$ (see Lemma \ref{lemma H}). Similarly, we also obtain
\begin{align}\label{step 1 b}
	& \| x (s_r+1)  - \hat x (s_r)\|_\infty  /\theta(s_r)   \nonumber\\
	& \le \| x (s_r+1) \! -\! \hat x (s_r)\|  /\theta(s_r)   = \|  e ^ x (s_r)  \!- \! \delta P (\eta(s_r))\|   /\theta(s_r)  \nonumber\\ 
	& = \|  e ^ x (s_r)  -  \delta P (\eta(s_r)) +  \delta P (x^*) \|   /\theta(s_r) \nonumber\\
	& = \|  e ^ x (s_r) \| /\theta(s_r)  +  \delta \|P (\eta(s_r)) -  \delta P (x^*) \|   /\theta(s_r) \nonumber\\
	& \le   \| \xi ^ x (s_r) \|   +   \delta  l \sqrt{N} \|x(s_r) -x^* \|   /\theta(s_r)    +   \delta   l  \|\bar y(s_r) \|   /\theta(s_r) \nonumber\\
	& \le   \| \xi ^ x (s_r) \|   +   \delta   l \sqrt{N}  \| \chi(s_r) \|      +   \delta   l  \| \beta (s_r) \|   \nonumber\\
	&\le N/2\gamma_1 + (1+\sqrt{N})   \delta   l C   \le  (2R+1)/2
\end{align}
in which the upper bounds of $\|\xi ^ x (s_r)\| $, $\|\beta(s_r)\| $ and $\|\chi(s_r)\|$ follow those in (\ref{step 1 a}).

By (\ref{step 1 a}) and (\ref{step 1 b}), it is clear that the quantizers for processing  $y_{ij}$ and $x_i$, respectively, are not saturated at $s_{r} + 1$.

\textbf{Step 2.} At $s_r + 2$, one needs to investigate the upper bounds of $\| y (s_r+2)  - \hat y (s_r+1)\|_\infty  /\theta(s_r+1) $ and $\| x (s_r+2)  - \hat x (s_r+1)\|_\infty  /\theta(s_r+1) $. Note that the analysis for $s_r + 2$ differs by the status of DoS attacks at $s_r + 1$, i.e., one needs to conduct the analysis for $s_r + 2$ by two cases: $s_r + 1$ is corrupted by DoS or not. 

2-1. If $s_r + 1$ is not corrupted by DoS, one can follow the very similar analysis in Step 1 to show the unsaturation of quantizers for $x_i$ and $y_{ij}$ at $s_r + 2$. 

2-2. If $s_r + 1$ is corrupted by DoS, then the quantizer for processing $y_{ij}$ at $s_r + 2$ is not saturated in view of 
\begin{align}\label{eq 43}
	&\| y (s_r+2)  - \hat y (s_r+1)\|_\infty   /\theta(s_r+1) \nonumber\\
	&\le \| y (s_r+1)  - \hat y (s_r+1)\|    /\theta(s_r+1)  \nonumber\\
	&= \| e^y (s_r + 1)\|    /\theta(s_r+1)  \nonumber\\
	&= \|	  G e^y (s_r) - h A_0 (\mathbf 1 _ N \otimes e^x (s_r)) - h S \bar y(s_r)  \|    /\theta(s_r + 1 )  \nonumber\\
	&\le \|G\| \|\xi ^ y (s_r)\| + h \|A_0\| \sqrt{N} \|\xi ^ x (s_r)\| + h \|S\| \|\beta(s_r)\|    \nonumber\\
	& \le  N\|G\|/2\gamma_1 \!+\! h \|A_0\|  N/2\gamma_1  \!+\! h \|S\| C    \le  (2R\!+\!1)/2.
\end{align}
In (\ref{eq 43}), we have used the fact $y(s_r +2) = y (s_r +1)$ in the first inequality, which is due to the presence of DoS at $s_r+1$ and hence $\sigma(k) = 0$ in (\ref{yij}b). 
Moreover, the quantizer for processing $x_{i}$ at $s_r + 2$ is not saturated by
\begin{align}
	& \| x (s_r+2)  - \hat x (s_r+1 )  \|_\infty  /\theta(s_r+1)   \nonumber\\
	& \le \| x (s_r+1)  - \hat x (s_r+1)\|  / \theta(s_r+1)  \nonumber\\
	& = \|  e ^ x (s_r + 1)   \|   /   \theta(s_r+1)    = \| e^x (s_r ) -   \delta P (\eta(s_r))      \|   /   \theta(s_r)    \nonumber\\ 
	& \le N/2\gamma_1+ (1+\sqrt{N})   \delta   l C    \le  (2R+1)/2
\end{align}
where we have used the fact $x(s_r +2) = x(s_r +1)$ in the first inequality because $s_r+1$ is corrupted by DoS and hence $u_i(k) = 0$ as shown in (\ref{u}).

\textbf{Step 3.} If $s_r + m -1$ ($m=2, 3, \cdots$) are all corrupted by DoS attacks consecutively, in view of Case 3 in (\ref{Case 3}), one can verify that the quantizers for processing $x_{i}$ and $y_{ij}$ are not saturated at $s_r + m $ since
$y (s_r+m)  - \hat y (s_r+ m- 1) = y (s_r+2)  - \hat y (s_r+1)$ and $ x (s_r+m)  - \hat x (s_r+m -1 ) =  x (s_r+2)  - \hat x (s_r +1 )$, in view of the analysis in Step 2.

\textbf{Step 4.} Suppose that $s_r + m $ in Step 3 is corrupted by DoS attacks and $s_r + m +1  \notin H_q$ (Case 4)). One can verify that the quantizers for quantizing $y_{ij}$ and $x_i$ are not saturated at $s_r + m +1$ by the following analysis, respectively:
\begin{align}\label{42}
	&\| y (s_r + m +1)  - \hat y (s_r + m   )\|_\infty   /\theta(s_r + m ) \nonumber\\
	&\le \| y (s_r + m)  - \hat y (s_r + m   )\|   /\theta(s_r + m ) \nonumber\\
	&= \| e^y (s_r + m)  \|   /\theta(s_r + m )    	= \| e^y (s_r + 1)  \|   /\theta(s_r + 1 ) \nonumber\\
	&= \| \xi^y (s_r + 1)  \|   \le (2R + 1)/2
\end{align}
and
\begin{align}\label{43}
	& \| x (s_r + m +1)  - \hat x (s_r + m  )  \|_\infty  /\theta(s_r+m)   \nonumber\\
	&\le \| x (s_r + m)  - \hat x (s_r + m  )  \|  /\theta(s_r+m)   \nonumber\\
	&= \| e ^ x (s_r + m)  \|  /\theta(s_r+m)     	= \| e ^ x (s_r + 1)  \|  /\theta(s_r+1)   \nonumber\\
	& =\|\xi ^ x (s_r + 1)\|     \le (2R + 1)/2 
\end{align}
where the first inequalities in (\ref{42}) and (\ref{43}) are obtained by Case 4 in Section III.

By the four steps above, one can see that the quantizers for processing $x_i$ and $y_{ij}$ are not overflowed at $s_r + 1$, $s_r + 2$, $\cdots$, $s_r + m $ and $s_r + m+1 = s_{r+1}$, which are essentially the steps during $(s_r, s_{r+1}]$. Namely, the transmissions between consecutive successful transmissions do not have quantizer saturation problem. By a similar analysis, one can verify that the quantizers processing $x_i$ and $y_{ij}$ during $[0, s_0]$ are not saturated if $0 \in H_q$. Overall, we conclude that the quantizers are not saturated at all $k$, i.e., result i) in Theorem \ref{theorem} and control objective 1).  

\textbf{Step 5.}
In the following, we prove result ii). By (\ref{transformation}) and (\ref{28}) (in Appendix-A), one can verify that $\|\pi(s_r+1)\|$ is upper bounded. Then by the first two equations in Cases 3) and 4), one can obtain $\|\pi(s_{r}+m)\| =
\|\pi(s_r+1)\|
$ for $m = 2, 3, \cdots$. Thus, one can see that $\|\pi(\cdot)\|$ is upper bounded at $s_r$ and also $s_r + m < s_{r+1}$ ($m=1, 2, \cdots$), which implies that $\|\pi(\cdot)\|$ is upper bounded at all $k$. By the transformation in (\ref{transformation}) and the scaling function in (\ref{eq h}), one can obtain
\begin{align}\label{47}
	\left\| \! 
	\begin{bmatrix} 
		\|   \bar y (k)  \| \\
		\| x(k) - x^*  \|
	\end{bmatrix}
	\! 
	\right\|
	& =  \theta(k)
	\|\pi(k)\| \nonumber\\
	&=
	\gamma_1 ^ {T_S (0, k)}  \gamma_2 ^ {T_U (0, k)} \theta(0)
	\|\pi(k)\|
\end{align} 
in which $T_U(0,k)$ denotes the number of unsuccessful transmissions between 0 and $k$. By $\gamma_2=1$, $\gamma_1 < 1$ and substituting $T_S(0,k)$ in Lemma \ref{Lemma T}, one has $\gamma_1 ^ {T_S (0, k)}\to 0$ as $k \to \infty$. Consequently, one has $
\left\|\begin{bmatrix} 
	\|   \bar y (k)  \| \\
	\| x(k) - x^*  \|
\end{bmatrix}\right\|
\to 0$ as $k \to \infty$, which implies $ y(k) \to  \mathbf 1_N \otimes x(k)$ and $x(k) \to x^*$ as $k \to \infty$, i.e., result ii) in Theorem \ref{theorem} and control objective 2).\qedp

\begin{remark}
	By Theorem \ref{theorem}, it is clear that if the amount of DoS attacks characterized by $1/T + \Delta/\tau_D$ is smaller than 1, the actions of players converge to the NE under quantized signals. As indicated in \cite{feng2017resilient}, this is ``the maximum resilience" one can achieve in the sense that if $1/T + \Delta/\tau_D \ge 1$, DoS attacks can be constantly present or fully synchronized with all the transmission attempts. If the communication links among the players are totally disabled, it is impossible to solve the distributed NE seeking problem for any controller, to the best of our knowledge. 
\end{remark}

\begin{remark}\label{remark 3}
	Our result regarding zooming-out in dynamic quantization is consistent with those of quantized consensus \cite{feng2023tcns} and quantized stabilization of centralized systems \cite{you2010minimum}. 	
	In \cite{feng2023tcns}, the agents have a general form (e.g., $x_i(k+1)=Ax_i(k) + Bu_i(k)$) and hence the zooming-out factor $\gamma_2$ in \cite{feng2023tcns} depends on the unstable eigenvalues of $A$.
	In our paper, the players take the form of an integrator or equivalently $A$, as a special case, is an identity matrix. Then the zooming out factor $\gamma_2$ takes the value 1. It is also possible to let $\gamma_2>1$. However, such a choice can reduce the convergence speed and importantly make the algorithm less robust against DoS attacks \cite{feng2023tcns}. Therefore, $\gamma_2=1$ is the optimal value corresponding to the player dynamics in this paper. If the players' dynamics follows $x_i(k+1)=Ax_i(k) + Bu_i(k)$ with $A$ having at least one unstable eigenvalues, one should let the zooming-out factor $\gamma_2>1$ to prevent quantizer saturation.  
\end{remark} 

\begin{remark}\label{remark 4}
	Convergence speed to the NE majorly depends on the choice of zooming-in parameter
	$\gamma_1$ and the amount of DoS attacks. 
	First, fix $\delta$ such that $\rho(\bar H)$ is minimized. 
	Subsequently, choose a small $\gamma_1$ satisfying $\rho(\bar H) < \gamma_1 <1 $. By (\ref{47}), one can find that a small $\gamma_1$ can increase the convergence speed. Note that a small $\gamma_1$ also implies a large data rate, i.e., a fast convergence speed requires more bandwidth. This is because a $\gamma_1$ close to $\rho(\bar H)$ can lead to a large $\gamma$ and hence a large $C$ (see (\ref{38}) and (\ref{44})). For the influence of DoS attacks, first note that when $1/T + \Delta /\tau_D$ approaches 0, $T_S(\cdot) $ in Lemma \ref{Lemma T} approaches $k$. Then $\gamma_1 ^ {T_S(\cdot)}$ converges to 0 quickly. This means that a mild DoS level admits a fast convergence speed and vice versa. Overall, a large network bandwidth and mild DoS attacks essentially lead to a large average data rate, which lead to a fast convergence speed. The discussion in this remark is also verified by the subsequent simulation results.  
\end{remark} 

\begin{figure}[t]
	\begin{center}
		\includegraphics[width=0.45 \textwidth]{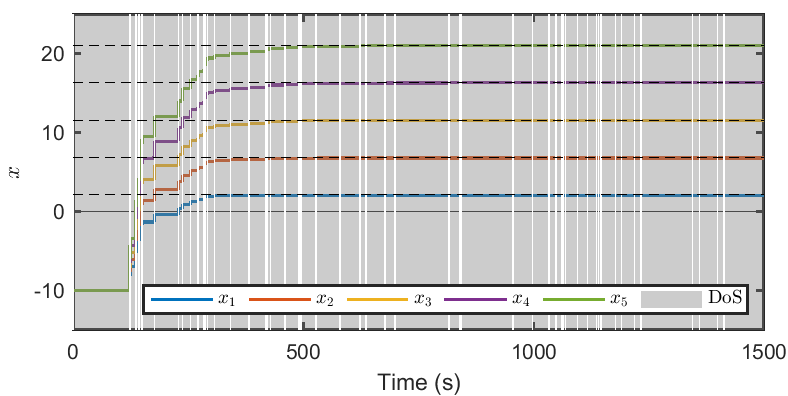}  \\
		\includegraphics[width=0.45 \textwidth]{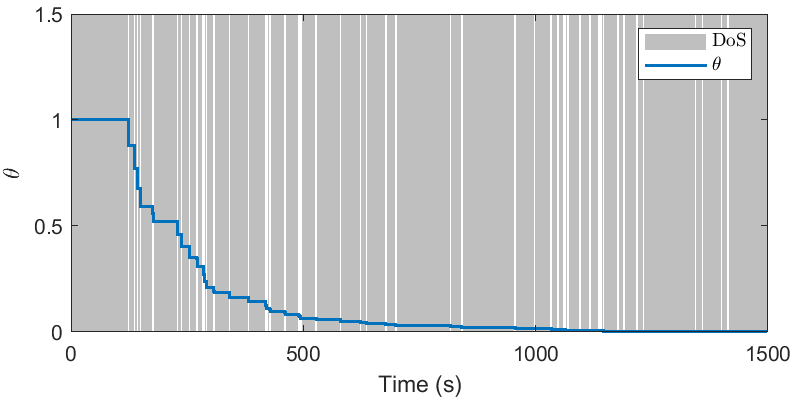}  \\
		\linespread{1}\caption{Time responses of $x(k)$ (top) and $\theta(k)$ (bottom). } \label{Fig1}
	\end{center}
\end{figure}

\begin{remark}\label{remark 5}
	We compare our results with those in relevant papers. Compared with the works of quantized consensus, our paper involves the nonlinear function $f_i(\cdot)$. Therefore, the method therein for selecting the zooming-in factor $\gamma_1$ is not applicable and new techniques in Lemma \ref{lemma H} are involved. Compared with \cite{chen2022distributed}, we consider the presence of DoS attacks. The techniques for analysis are also different, i.e., the analysis in \cite{chen2022distributed} is based on a Lyapunov approach and ours is aligned with quantized consensus in \cite{li2010distributed, you2011network, feng2023tcns}. 
	Compared with \cite{ye2017distributed}, our paper additionally considers the presence of DoS attacks and limited communication bandwidth. Both our paper and \cite{9878291}  consider quantized distributed NE seeking problems. As significant improvements, the control architecture in our paper can achieve asymptotic convergence to the NE instead of a practical convergence in \cite{9878291}. Moreover, the ranges of quantizers preventing saturation are explicitly provided in our paper, which are not provided in \cite{9878291}. 
\end{remark}

\begin{figure}[t]
	\begin{center}
		\includegraphics[width=0.45 \textwidth]{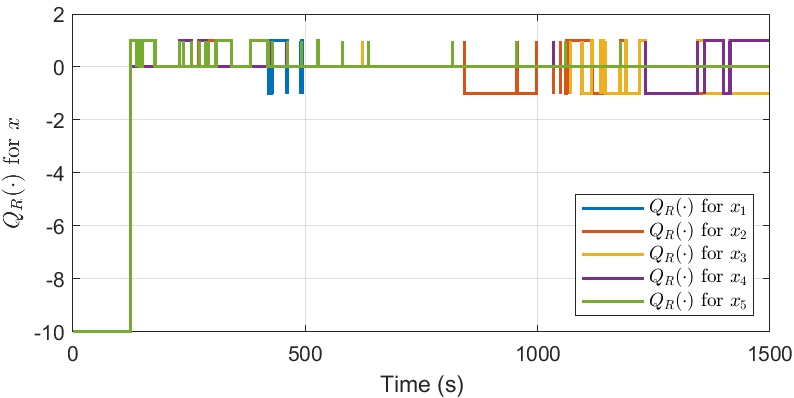}  \\
		\includegraphics[width=0.45 \textwidth]{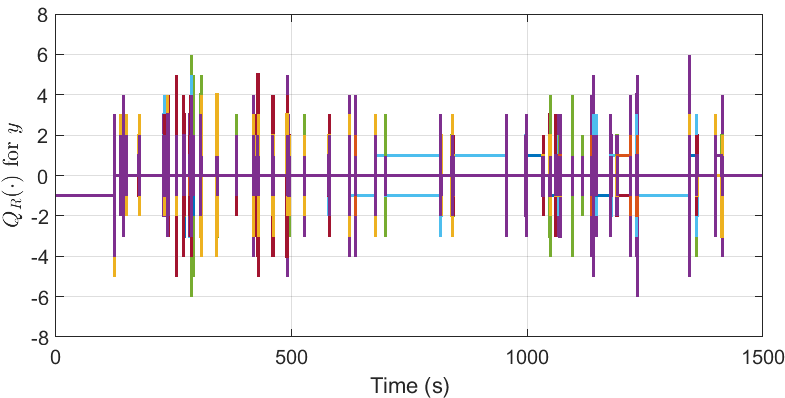}  \\
		\linespread{1}\caption{The values of $Q_R(\cdot)$ for $x$ and $y$. The legend for the bottom plot is similar to that in the top plot and hence is omitted.} \label{Fig3}
	\end{center}
\end{figure}

\section{Simulation}

In this section, we consider the payoff function $f_i(x)=\rho_i(x_i-x_i^d)^2 +(p_0\sum_{i=1}^{N}x_i+q_0)x_i$ for player $i$ in \cite{ye2017distributed} with $i = 1, 2, 3, 4, 5$, which can be used to model the game of energy consumption for heating ventilation and air conditioning systems. The values of $\rho_i, x_i^d, p_0, q_0$ and $x(0)$ in $f_i(x)$ follow those in \cite{ye2017distributed}, under which the unique NE is $x^*=[2.0147\,\,6.7766\,\, 11.5385\,\, 16.3004\,\, 21.0623]^T$. The communication topology among the players follows that in \cite{chen2022distributed} and select $\Delta=0.01$s for periodic transmissions. 

We choose $h=0.44$ such that $\|H\|=0.8571$ by Lemma \ref{lemma H}. Then we choose $\delta=0.001$ such that $\rho(\bar H)=0.9991$ by Appendix-C. By Lemma \ref{lemma H} and Theorem \ref{theorem}, we choose $\gamma_1=0.9992$ and obtain the conditions of $R$ for quantizer unsaturation: 
\begin{subequations}\label{49}
	\begin{align}
		&\text{For $x_i$: }  \frac{2R + 1}{2} \ge  851 \\
		&\text{For $y_{ij}$: }  \frac{2R + 1}{2} \ge  179960. 
	\end{align}
\end{subequations}

For DoS attacks, we consider a 
sustained DoS attack with variable period and duty cycle, generated randomly.
Over a simulation horizon of $1500$s, the DoS 
signal yields $|\Xi(0,1500)|=1465$s and $n(0,1500)=50$. 
This corresponds to values (averaged over $1500$s) 
of $\tau_D\approx 30$ and $1/T \approx 0.9767$,
and hence
$
\Delta/\tau_D + 1/T \approx 0.98
$, which approaches the maximum tolerable DoS attack in Theorem \ref{theorem}. In Fig. \ref{Fig1}, most of the simulation time is occupied by gray stripes implying the presence of DoS. The white stripes represent DoS-free intervals. Under such a strong level of DoS attacks, one can see in the top plot of Fig. \ref{Fig1} that the actions of the players converge to the NE. In the bottom plot in Fig. \ref{Fig1}, one can see that the scaling function $\theta$ converges to zero as time elapses. Moreover, $\theta$ holds its value in the presence of DoS and decreases during DoS-free intervals, i.e., the zooming-in and holding. In Fig. \ref{Fig3}, we present the quantized values for $x$ and $y$, i.e., (\ref{hat xj}) and (\ref{hat yij}), respectively. One can see that the actual utilized quantization range of $x$ is between $-10$ and $1$, and the one for $y$ is between $-6$ and $6$. They do not encounter saturation problem under DoS attacks and both satisfy the conditions for quantizer unsaturation in (\ref{49}). One can see that the result in (\ref{49}b) is conservative. One of the major reasons is the multiplication of $\|G\|$, $\|A_0\|$ and $\|S\|$ in (\ref{result theorem}b) for calculating the value in (\ref{49}b). Note that constructiveness of quantization levels often occurs in quantized control of multi-agent systems even without DoS attacks (see e.g., \cite{you2011network, li2010distributed}). This is majorly due to the nature of distributed systems. If global state is unknown, the agents cannot ``well" estimate the states of their neighbors. By contrast, for centralized systems, the problem of constructiveness can be well tackled \cite{you2010minimum}.

\begin{figure}[t]
	\begin{center}
		\includegraphics[width=0.45 \textwidth]{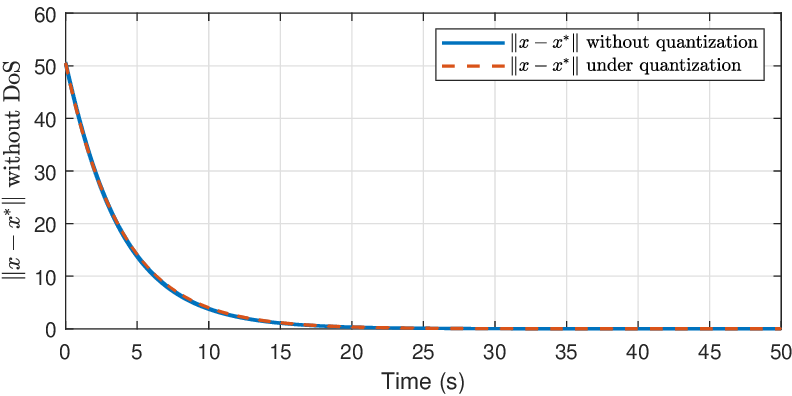}  \\
		\linespread{1}\caption{Dynamics of $\|x(k)-x^*\|$ without and under quantization.} \label{Fig4}
	\end{center}
\end{figure}

At last, we compare the dynamics of $x$ for the cases between quantized and unquantized NE seeking in Fig \ref{Fig4}. One can see that the convergence speeds are very close and the case of quantization-free has a slightly faster convergence speed. This is because we select a $\gamma_1=0.9992$ very close to $\rho(\bar H)=0.9991$. However, as mentioned in Remark \ref{remark 4}, this also causes large data rates as shown in (\ref{49}).

In the above simulations, we have shown in Fig. \ref{Fig3} that the actual utilized quantization range of $x$ is $[-10,1]$, and the one for $y$ is $[-6,6]$, under which the players' state can reach the NE. 
We mention that the players' actions can still reach the NE under the quantization range $[-1,1]$, i.e., $R=1$ in (\ref{quantizer}). This implies that our algorithm is quite robust in the sense that even if the simulation results in Fig. \ref{Fig3} are violated, one can still fulfill the NE seeking. Eventually, the players fail to reach the NE if the quantizer follows $q_R(b)=0$ if $-0.5<b<0.5$ and $q_R(b)=\text{sign}(b) 0.5$ if $|b|\ge 0.5$.


\section{Conclusions and future research}


This paper proposed a quantized distributed NE seeking algorithm under DoS attacks. To mitigate the influence of DoS attacks, a zooming-in and holding quantization scheme was developed. A sufficient condition on the numbers of quantization levels was provided, under which the quantizers do not saturate under DoS attacks. Our quantized distributed NE seeking strategy has the maximum resilience to DoS attacks, namely, beyond the bound of the maximum resilience, all the transmission attempts can be denied and distributed NE seeking is impossible.

In the future, it is interesting to explore quantized distributed NE seeking under DoS attacks by Lyapunov-based approaches by following \cite{chen2022distributed}. Extending our results to the players in a general linear form is also a practical direction\cite{feng2023tcns}. It is also useful to apply our results to defend cyber-attacks \cite{zhu2015game}. At last, we expect that the results in our paper still hold under a directed and strongly connected graph in view of \cite{ye2017distributed}.

\appendix
\subsection{Proof of $\|\xi^x(s_r)\|_\infty \le 1/2\gamma_1$ and $\|\xi^y(s_r)\|_\infty \le 1/2\gamma_1$ }
By Case 1) with $k+1 \notin H_q$, one can obtain 
\begin{subequations}\label{21}
	\begin{align}
		& \xi ^ x (k+1) =  \frac{\xi^x (k) -  \delta P (\eta(k))/\theta(k)}{\gamma_1}  \nonumber\\
		& \quad \quad \quad \quad\quad   - \frac{Q _R \left(  \xi^x (k) -  \delta P (\eta(k)) /\theta(k)  \right)}{\gamma_1} \\
		& \xi^ y (k+1) = \frac{G \xi^y (k) - h A_0 (\mathbf 1 _ N \otimes \xi^x (k)) - h S \beta(k) }{\gamma_1}\nonumber\\
		&\quad\quad   - \frac{ Q_R \left(G \xi^y (k) \!-\! h A_0 (\mathbf 1 _ N \otimes \xi^x (k)) \!-\! hS \beta(k)  \right)}{\gamma_1}.
	\end{align}
\end{subequations}
Similarly, for Case 4) with $k+1 \notin H_q$, one has 
\begin{subequations}\label{22}
	\begin{align}
		\xi^ x (k+1) &=  \frac{\xi^x (k)  -   Q_R \left( \xi^x(k)\right)}{\gamma_1}  \\
		\xi^ y (k+1) & =   \frac{\xi^y (k)   -  Q_R \left( \xi^ y(k)\right)}{\gamma_1}.
	\end{align}
\end{subequations}
In (\ref{21}) and (\ref{22}), one can verify that $\|\xi^x(k+1)\|_\infty \le 1/2\gamma_1$ and $\|\xi^y(k+1)\|_\infty \le 1/2\gamma_1$ if $Q_R(\cdot)$ does not saturate. Since $k+1 \notin H_q$, namely $s_r$, then one has $\|\xi^x(s_r)\|_\infty \le 1/2\gamma_1$ and $\|\xi^y(s_r)\|_\infty \le 1/2\gamma_1$ 
if $Q_R(\cdot)$ does not saturate. \qedp

\subsection{Proof of Lemma \ref{lemma H}}
We conduct the analysis in the following order: $s_r=k \notin H_q $, $\{k+1, k+2, \cdots, k+m   \} \in H_q $ and $s_{r+1}=k+m +1 \notin H_q $, in order to investigate the system dynamics between two consecutive successful transmissions. 

\textbf{Step 1.} For $s_r = k \notin H_q $ and $k+1 \in H_q$ (corresponding to Case 2)), by (\ref{Case 2}a), one can obtain 
\begin{align}\label{bary case2}
	&\| \bar y (k+1)\| \nonumber\\
	&\le \| H \| \|\bar y (k)\| + \delta  \sqrt{N} \|P (\eta(k))\|  \nonumber\\
	&\quad 	+ h \|S\| \|e^y(k)\| + h \|A_0\| \sqrt{N} \|e ^ x (k)\| \nonumber\\
	& \le   \| H \| \|\bar y (k)\|  + \delta   \sqrt{N}  \|P (\eta(k)) - P (x(k)) +   \nonumber\\
	&\quad P(x(k))-  P (x ^*)\| + h \|S\| \|e^y(k)\| + h \|A_0\| \sqrt{N} \|e ^ x (k)\| \nonumber\\
	& \le   \| H \| \|\bar y (k)\|  + \delta   N  l  \| x(k) - x ^*\|  +   \delta   \sqrt{N}   l  \| \bar y(k) \|  \nonumber\\
	&\quad 	+ h \|S\| \|e^y(k)\| + h \|A_0\| \sqrt{N} \|e ^ x (k)\|  \nonumber\\
	&=  ( \| H \| +  \delta \sqrt{N}  l ) \|\bar y (k)\|  + \delta  N   l  \| x(k) - x ^*\|    \nonumber\\
	&\quad 	+ h \|S\| \|e^y(k)\| + h \|A_0\| \sqrt{N} \|e ^ x (k)\|.  
\end{align}
where we have used $ \|P (\eta(k)) - P (x(k)) + P(x(k))-  P (x ^*)\| \le   l \|y(k) - \mathbf 1 _ N \otimes  x(k)\|  +   l \sqrt{N} \|x(k) - x^*\|$.

By (\ref{Case 2}b), one has
\begin{align}\label{x case2}
	&\| x(k+1) - x^* \| \nonumber\\
	& = \|  x(k) - x^* - \delta   P(\eta(k)) \|   \nonumber\\
	& =\|    x(k) - x^* - \delta  P(x(k)) +   \delta  P(x(k))- \delta   P(\eta(k))  \| \nonumber\\
	&\le   \|    x(k) - x^* - \delta   P(x(k)) \| +   \delta  \|P(x(k))-  P(\eta(k))  \| \nonumber\\
	& \le  \sqrt{1-2\delta   \mu + \delta ^2  l ^2} \|    x(k) - x^*  \| +  N \delta    l \| \bar y(k) \|  
\end{align}
where the first term in the last inequality is obtained by 
\begin{align}\label{25}
	&\|x(k) - x^* - \delta  P(x(k))\|^2  \nonumber\\
	&=\|x(k)-x^*\|^2 - 2  \delta  (x(k)-x^*) ^T P(x(k)) +  \delta  ^2 \|P(x(k))\|^2 \nonumber\\
	&= \|x(k)-x^*\|^2 - 2 \delta (x(k)-x^*) ^T ( P(x(k)) - P (x^*))\nonumber\\
	&\quad+  \delta  ^2 \|P(x(k))- P (x^*)\|^2 \nonumber\\
	& \le \|x(k)-x^*\|^2 - 2 \delta  \mu \|x(k)-x^*\|^2 +  \delta  ^2    l ^ 2 \|x(k)- x^*\|^2 \nonumber\\
	&=(1- 2 \delta   \mu + (\delta    l)^2)\|x(k)-x^*\|^2
\end{align}
and the second term in the last inequality of (\ref{x case2}) is obtained by $ \|P(x(k))-  P(\eta(k))  \| \le   l \|  y(k)  - \mathbf 1 _ N \otimes x(k)\|=   l \|  \bar  y(k) \| \le N l \|\bar y (k)\|$. 
In (\ref{25}), one can verify that $1- 2 \delta   \mu + (\delta     l)^2$ is no smaller than 0 by the fact
$
1- 2 \delta   \mu + (\delta    l)^2 = (  l \delta   -1)^2 + 2 (  l - \mu )\delta   \ge 0
$
if $  l \ge \mu$, and 
$
1- 2 \delta \alpha \mu + (\delta \alpha   l)^2 \le	1- 2 \delta \alpha \mu + (\delta \alpha \mu)^2   
= (1-\mu\delta\alpha)^2 
\ge 0
$
if $\mu \ge   l$. Moreover, in the first inequality of (\ref{25}), we have used $(x(k)- x^*) ^T (P(x(k)) - P(x^*)) \ge \mu \| x(k) - x^*\|^2$ implied by Assumption \ref{ass 2}.

By collecting (\ref{bary case2}) and (\ref{x case2}) into a compact form, we obtain 
\begin{align}\label{compact}
	&\begin{bmatrix} 
		\| \bar y (k+1)  \| \\
		\|x(k+1) - x ^ * \|
	\end{bmatrix}
	\le 
	\bar H
	\begin{bmatrix}
		\| \bar y (k)  \| \\
		\|x(k) - x ^ * \|
	\end{bmatrix} \nonumber\\
	&\quad\quad + 
	\underbrace{\begin{bmatrix}
			h \|S\| \\
			0
	\end{bmatrix}}
	_{\bar C}
	\|e ^y (k)\|
	+ 
	\underbrace{\begin{bmatrix}
			h \|A_0\| \sqrt{N} \\
			0
	\end{bmatrix}}
	_{\bar D}
	\|e ^ x (k)\|. 
\end{align}
Note that there exists a positive $ \delta$ such that $\rho(\bar H)<1$, whose proof is provided in Appendix-C. Since $k \notin H_q$ can be denoted by $s_r$, then (\ref{compact}) can be rewritten as  
\begin{align}\label{28}
	\begin{bmatrix} 
		\| \bar y (s_r+1)  \| \\
		\|x(s_r+1) - x ^ * \|
	\end{bmatrix}
	&\le 
	\bar H 
	\begin{bmatrix}
		\| \bar y (s_r)  \| \\
		\|x(s_r) - x ^ * \|
	\end{bmatrix} 
	\nonumber\\
	&\quad + 
	\bar C
	\|e ^y (s_r)\|
	+ 
	\bar D
	\|e ^ x (s_r)\|
\end{align}
where $s_r + 1 = k+1$ is a DoS corrupted step by hypothesis, i.e., $s_r + 1 \in H_q$.

\textbf{Step 2.}
Now we consider the state evolution between $s_r + 1  \in H_q $ and $s_r + 2  \in H_q$, namely Case 3). One can verify that $\bar y(s_r + 2) = \bar y (s_r + 1)$ and $ x(s_r + 2 ) - x^* = x (s_r  + 1) - x^* $. Similarly, if the transmissions at $s_r + m $ ($m = 3, 4, \cdots)$ consecutively fail due to DoS attacks, then $\bar y(s_r + m) =\bar y (s_r + m-1)=\cdots = \bar y (s_r + 1)$ and $ x(s_r + m ) - x^* = x(s_r + m -1 ) - x^* = \cdots = x (s_r  + 1) - x^* $.

\textbf{Step 3.}
Now we consider the state evolution between $s_r + m  \in H_q $ and $s_r + m +1 \notin H_q$, namely Case 4). Accordingly, we have $\bar y(s_r + m+ 1) = \bar y (s_r + m)$ and $ x(s_r + m +1) - x^* = x (s_r  + m) - x^* $, in which $\bar y (s_r + m)$ and $ x(s_r + m +1) - x^* $ were obtained in Step 2.

\textbf{Step 4.} By the analysis in Steps 1-3, for $s_r=k  \notin H_q $,  $\{k+1, k+2, \cdots, k+m   \} \in H_q $  and $s_r+m +1 \notin H_q $, we have 
\begin{align}\label{34}
	\begin{bmatrix} 
		\| \bar y (s_r+m+1)  \| \\
		\|x(s_r\!+\!m\!+\!1) - x ^ * \|
	\end{bmatrix} 
	&\le 
	\bar H 
	\begin{bmatrix}
		\| \bar y (s_r)  \| \\
		\|x(s_r) - x ^ * \|
	\end{bmatrix} 
	\nonumber\\ 
	&\quad + 
	\bar C
	\|e ^y (s_r)\|
	+ 
	\bar D
	\|e ^ x (s_r)\|. 
\end{align}
Since $s_r+m +1 \notin H_q $ can be denoted by $s_{r+1} $, then (\ref{34}) can be written as 
\begin{align}\label{compact 1}
	\begin{bmatrix} 
		\| \bar y (s_{r+1})  \| \\
		\|x(s_{r+1}) - x ^ * \|
	\end{bmatrix}
	& \le 
	\bar H 
	\begin{bmatrix}
		\| \bar y (s_r)  \| \\
		\|x(s_r) - x ^ * \|
	\end{bmatrix} 
	\nonumber\\ 
	&\quad + 
	\bar C
	\|e ^y (s_r)\|
	+ 
	\bar D
	\|e ^ x (s_r)\|. 
\end{align}
Therefore, \eqref{compact 1} describes the system dynamics between consecutive successful transmissions at $s_r$ and $s_{r+1}$, between which packet losses can occur due to DoS attacks.

\textbf{Step 5.} In view of (\ref{compact 1}), by defining $e(s_r):= \bar C \| e^y(s_r)\| + \bar D \|e ^x (s_r)\| \in \mathbb R ^2$, one has 
\begin{align}\label{33}
	&	\frac{	\begin{bmatrix} 
			\|   \bar y (s_{r+1})  \| \\
			\| x(s_{r+1}) - x^*  \|
	\end{bmatrix} }{\theta(s_r)}
	\le 
	\bar H
	\frac{\begin{bmatrix}
			\| \bar y  (s_r)  \| \\
			\| x(s_r) - x^* \|
	\end{bmatrix}}{\theta(s_r)}  + 
	\frac{e(s_r)}{\theta(s_r)} \Rightarrow \nonumber\\ 
	&	\frac{	\begin{bmatrix} 
			\|   \bar y (s_{r+1})  \| \\
			\| x(s_{r+1}) - x^*  \|
	\end{bmatrix} }{\theta(s_{r+1})/\gamma_1}
	\le 
	\bar H
	\frac{\begin{bmatrix}
			\| \bar y  (s_r)  \| \\
			\| x(s_r) - x^* \|
	\end{bmatrix}}{\theta(s_r)}  + 
	\frac{e(s_r)}{\theta(s_r)} 
\end{align}
in which we have utilized the fact $\theta(s_r) = \theta(s_{r+1}) / \gamma_1$ due to the design of scaling function (\ref{eq h}), i.e., $\theta(k)$ zooms in by $\gamma_1$ for successful transmissions and holds the value by $\gamma_2=1$ under DoS attacks. 
By the transformations in (\ref{transformation}) and the definition of $\pi(k)$ in (\ref{pi}), inequality (\ref{33}) can be transformed into
\begin{align}\label{compact 2}
	\!\!\!\!
	\pi(s_r+1)
	& \!\!\le\! 
	\frac{\bar H}{\gamma_1}
	\pi(s_r)  + 
	\frac{1}{\gamma_1}
	\xi(s_r)\nonumber\\
	&\!\!\le \!\!
	\left(\!\frac{\bar H}{\gamma_1}\!\right) ^ {r+1} \!\!
\pi(s_0)\!\! +\!\! \frac{1}{\gamma_1}\! \sum_{i = 0}^{r}\!\left(\!
	\frac{\bar H}{\gamma_1}\!\right)^{r-i} \!\!\!\xi(s_i).
\end{align}

In  (\ref{compact 2}), $\xi(s_i)$ is bounded for $i = 1, \cdots, r$.  Specifically, one has $\xi(s_i) := e(s_i)/\theta(s_i)=\bar	C \|\xi ^y (s_i)\|+ \bar D\|\xi ^ x (s_i)\| \in \mathbb R ^{2}_{\ge 0}$ with $\|\xi ^ y (s_i)\| \le N \|\xi ^ y (s_i)\| _ \infty  \le \frac{N}{2\gamma_1}$ and $\|\xi ^ x (s_i)\| \le \sqrt{N} \|\xi ^ x (s_i)\| _ \infty  \le \frac{\sqrt{N}}{2\gamma_1}$ due to the hypothesis in the lemma. That is, the quantizer is not saturated at $s_i$ for $i=1, \cdots, r$, hence we have $\|\xi ^ x (s_r)\| _ \infty \le 1/2\gamma_1 $ and $\|\xi ^ y (s_r)\| _ \infty \le 1/2\gamma_1 $. Substituting $\bar C$ and $\bar D$, we have $\|\xi(s_r)\| \le  \frac{h N}{2\gamma_1}(\|S\|+\|A_0\|)$. Note that the bound above also holds for $r=-1$.

In the following, we show that $\|\pi(s_0)\|$ in (\ref{compact 2}) is upper bounded by $\sqrt{N^2 C_{x_0} ^2 + N (C_{x^*} + C_{x_0})^2}/\theta_0$. By $\pi(k)=	 
[	\|   \beta (k)  \| \,\,
	\|\chi (k)  \| ]^T$, it is sufficient to calculate the upper bounds of $\|\beta(s_0)\|$ and $\|\chi (s_0)\|$ individually. We present the analysis by the following two cases. If $s_0 = 0$ (i.e., there has no DoS attack at the beginning), we have $
\|\beta (s_0)\| = \| \bar y (s_0)\| / \theta(s_0) = \|\bar y (0)\| /\theta(0)= \|y(0) - \mathbf 1 _N \otimes x(0)\| / \theta(0)= \| \mathbf 1 _N \otimes x(0)\| / \theta(0)  \le N  \| \mathbf 1 _N \otimes x(0)\|_\infty / \theta(0) = N C_{x_0}/ \theta(0)$. Similarly, one has $\|\chi (s_0)\|  = \|x(s_0) - x^*\| /\theta(s_0) \le (\|x(s_0)\| + \|x^*\|)  / \theta(0) \le \sqrt{N}(\|x(s_0)\|_\infty + \|x^*\| _\infty) /\theta(0) \le \sqrt{N} (C_{x^*} + C_{x_0})/\theta(0)$. If $s_0 > 0$ (i.e., there are DoS attacks since the initial time), by the analysis in Steps 1-3, one can verify that $\bar y(s_0) = \bar y(0)$, $x(s_0) = x(0)$ and $\theta(s_0)=\theta(0)$. This implies that $\|\pi(s_0)\| = \|\pi(0)\|$, whose value has been provided in the case of $s_0 =0$. Overall, for both cases, one has  $\|\pi(s_{-1})\|=\|\pi(s_0)\|\le \sqrt{N^2 C_{x_0} ^2 + N (C_{x^*} + C_{x_0})^2}/\theta_0$.

As we have selected $\rho(\bar H)<\gamma_1<1$, then there exist $C_ \gamma \in \mathbb R _{\ge 1}$ and $\rho(\bar H)/ \gamma_1 \le \gamma<1$ such that $\| (\bar H  /\gamma_1)^k \|\le C_ \gamma \gamma ^ k$. By \eqref{compact 2}, one has  
\begin{align}\label{38}
\|\pi(s_r)\| &\le \gamma^r \frac{\gamma C_ \gamma}{\theta_0}\sqrt{N^2 C_{x_0} ^2 + N (C_{x^*} + C_{x_0})^2}
	\nonumber\\
	&\quad + (1-\gamma^r) \frac{C_{\gamma}hN(\|S\|+\|A_0\|)}{2\gamma_1^2(1-\gamma)} \nonumber\\
	& \le  \max\left\{\frac{\gamma C_{\gamma} }{\theta_0}\sqrt{N^2 C_{x_0} ^2 + N (C_{x^*}+ C_{x_0})^2}, \right. \nonumber\\
	&\quad\quad\quad\quad\left.\frac{C_{\gamma}hN(\|S\|+\|A_0\|)}{2\gamma_1^2(1-\gamma)} \right\}.
\end{align}
Incorporating the cases of $s_{-1}$ (if $s_{0} \ne 0$) and $s_0$ (if $s_{0}= 0$), we have 
\begin{align}\label{44}
	\|\pi(s_r)\|
	& \le  \max\left\{\frac{\bar C_\gamma}{\theta_0}\sqrt{N^2 C_{x_0} ^2 + N (C_{x^*}+ C_{x_0})^2}, \right. \nonumber\\
	&\quad\quad\quad\quad\left.\frac{C_{\gamma}hN(\|S\|+\|A_0\|)}{2\gamma_1^2(1-\gamma)} \right\}=:C
\end{align}
in which $\bar C_\gamma:=\max\{\gamma C_{\gamma} ,1\}$ and $r=-1, 0, \cdots$.
\qedp

\subsection{Proof for the existence of $\delta$}

We will show that there exists a positive $\delta $ such that $\rho(\bar H ) < 1$. To verify this, it is sufficient to check if there exists a $\delta$ simultaneously satisfying \cite{xiong2022quantized}: 
\begin{subequations}\label{eq conditions}
	\begin{align}
		&1- 2 \delta  \mu + (\delta  l)^2 <1  \Rightarrow \delta  < \frac{2\mu}{  l^2} \\
		&	\|H\| + \delta    l \sqrt{N} < 1 \Rightarrow \delta  < \frac{1- \|H \|  }{\mu \sqrt{N}} \\
		& \text{det}( I_2 - \bar H )  > 0   \Rightarrow  B(\delta):=(1-\|H\|-    l \sqrt N \delta ) \times \nonumber\\
		&\quad\quad\left(1-\sqrt{1-2\mu\delta +  l^2 (\delta )^2}\right)
		- N^2   l^2  (\delta)^2>0.  
	\end{align}
\end{subequations}
It is clear that a sufficiently small $\delta$ can satisfy (\ref{eq conditions}a) and (\ref{eq conditions}b). Note that $1-\|H\|>0$ in (\ref{eq conditions}b) since $\|H\| = \rho(H)<1$ implied by Lemma \ref{lemma: h}. In the following, we show the existence of $\delta $ satisfying (\ref{eq conditions}c). We compute the derivative of $B(\delta)$ along $\delta$ and obtain
\begin{align}
	D(\delta) &:=  \frac{ dB(\delta)}{d\delta} =-  l \sqrt{N} (1-\sqrt{1-2\mu\delta +  l ^2 \delta ^2})  \nonumber\\
	&\quad - 2N^2  l ^2 \delta  -(1-\|H\|-   l \sqrt N \delta ) \times \nonumber\\
	&\quad ((1-2\mu\delta +  l ^2 \delta  ^2)(2  l^2 ( \delta^2-2\mu)).
\end{align} 
First, it is simple to verify that $D(0)=2\mu(1-\|H\|)>0$ and $B(0)=0$. Since $D(\delta )$ is continuous and differentiable for $0 \le \delta < \min \{2\mu/  l ^2, (1-\|H\|)/ \mu \sqrt N\}$, this implies that there exists a $0<\delta^* <\min \{2\mu/  l ^2, (1-\|H\|)/ \mu \sqrt N\}$ such that $D(\delta )  >0$ for $0< \delta <      \delta^*  $ and hence $B(\delta)>0$. \qedp

\bibliographystyle{IEEEtran}

\bibliography{ref}

\end{document}